\documentclass[sigconf]{acmart}

\usepackage{graphicx} % Required for inserting images
\usepackage{xcolor}
\usepackage{hyperref}
\usepackage{todonotes}
\usepackage{multirow}
\usepackage{hhline}
\usepackage{amsmath}
\usepackage{amsthm}
\usepackage{amsfonts}
\usepackage{algorithm}
\usepackage{algpseudocode}
\usepackage{bbm}
\usepackage{caption}
\usepackage{subcaption}
\usepackage{bbm}
\usepackage{amsmath}
\usepackage{tikz}
\usepackage{adjustbox}
\usepackage{xspace}
\usepackage{flushend}
\usepackage{bm}

\newtheorem{problem}{Problem}

\usetikzlibrary{positioning}
\usetikzlibrary{shapes.geometric}
\usetikzlibrary{shapes,arrows,chains}
\usetikzlibrary{arrows.meta}
\usetikzlibrary[calc]

\newcommand{\aris}[1]{\textcolor{orange}{\textbf{Aris}: {#1}}\xspace}
\newcommand{\erica}[1]{\textcolor{blue}{\textbf{Erica}: {#1}}\xspace}

\DeclareMathOperator*{\argmax}{argmax} 

\algnewcommand{\algorithmicinput}{\textbf{Input}}

\newcommand{\para}[1]{\noindent{\bf #1}}
\newcommand{\spara}[1]{\smallskip\noindent{\bf #1}}

\newcommand{\vect}[1]{\ensuremath{\mathbf{#1}}\xspace}

\newcommand{\expectation}[1]{\ensuremath{\mathbb{E}\!\left[{#1}\right]}\xspace}

\newcommand{\reals}{\ensuremath{\mathbb{R}}\xspace}
\newcommand{\users}{\ensuremath{\mathcal{U}}\xspace}
\newcommand{\auser}{\ensuremath{u}\xspace}
\newcommand{\nousers}{\ensuremath{m}\xspace}
\newcommand{\items}{\ensuremath{\mathcal{I}}\xspace}
\newcommand{\itemset}{\ensuremath{\mathcal{X}}\xspace}
\newcommand{\anitem}{\ensuremath{i}\xspace}
\newcommand{\anotheritem}{\ensuremath{j}\xspace}
\newcommand{\noitems}{\ensuremath{n}\xspace}
\newcommand{\strategy}{\ensuremath{\mathcal{S}}\xspace}

\newcommand{\score}{\ensuremath{\mathcal{P}}\xspace}
\newcommand{\normscore}{\ensuremath{\mathcal{\widehat{P}}}\xspace}
\newcommand{\relscore}{\ensuremath{\mathcal{R}}\xspace}
\newcommand{\combinedscore}{\ensuremath{\mathcal{Z}}\xspace}
\newcommand{\normrelscore}{\ensuremath{\mathcal{\widehat{R}}}\xspace}
\newcommand{\idiversity}{\ensuremath{\mathcal{T}}\xspace}
\newcommand{\normsumdiversity}{\ensuremath{\mathcal{\widehat{T}}}\xspace}
\newcommand{\sumdistance}{\ensuremath{\mathcal{D}}\xspace}

\newcommand{\sumcoverage}{\ensuremath{\mathcal{C}}\xspace}

\newcommand{\vectx}{\ensuremath{\vect{x}}\xspace}
\newcommand{\vectxi}{\ensuremath{{\vectx}_{\anitem}}\xspace}
\newcommand{\vectxiu}{\ensuremath{{x}_{\anitem\auser}}\xspace}
\newcommand{\categories}{\ensuremath{\mathcal{C}}\xspace}
\newcommand{\acategory}{\ensuremath{c}\xspace}
\newcommand{\vecty}{\ensuremath{\vect{y}}\xspace}
\newcommand{\vectyi}{\ensuremath{{\vecty}_{\anitem}}\xspace}
\newcommand{\vectyic}{\ensuremath{{y}_{\anitem\acategory}}\xspace}
\newcommand{\distance}{\ensuremath{d}\xspace}
\newcommand{\setw}{\ensuremath{\mathcal{W}}\xspace}
\newcommand{\vectz}{\ensuremath{\vect{z}}\xspace}
\newcommand{\vectzi}{\ensuremath{{\vectz}_{\anitem}}\xspace}
\newcommand{\steps}{\ensuremath{\mathrm{\kappa}}\xspace}
\newcommand{\diversity}{\ensuremath{\mathrm{div}}\xspace}
\newcommand{\divcov}{\ensuremath{\diversity_{\!_C}}\xspace}
\newcommand{\divdist}{\ensuremath{\diversity_{\!_D}}\xspace}

\newcommand{\ilist}{\ensuremath{\mathbf{L}}\xspace}
\newcommand{\iter}{\ensuremath{t}\xspace}
\newcommand{\ilistk}{\ensuremath{\ilist_{\iter}}\xspace}
\newcommand{\ilistsize}{\ensuremath{k}\xspace}
\newcommand{\seenitems}{\ensuremath{\mathcal{X}}\xspace}
\newcommand{\seenitemsk}{\ensuremath{\mathcal{X}_{\iter}}\xspace}
\newcommand{\candidates}{\ensuremath{\mathcal{J}}\xspace}
\newcommand{\candidatesk}{\ensuremath{\candidates_{\iter}}\xspace}

\newcommand{\selectprob}{\ensuremath{p}\xspace}
\newcommand{\selectprobi}{\ensuremath{\selectprob_{\anitem}}\xspace}

\newcommand{\bufferprob}{\ensuremath{q}\xspace}
\newcommand{\bufferprobi}{\ensuremath{\bufferprob_{\anitem}}\xspace}
\newcommand{\quitatlistk}{\ensuremath{Q_{\iter}}\xspace}
\newcommand{\totalquit}{\ensuremath{Q_T}\xspace}

\newcommand{\quitprob}{\ensuremath{\eta}\xspace}
\newcommand{\quitprobk}{\ensuremath{{\quitprob}_{\iter}}\xspace}

\newcommand{\ouralgo}{\textsc{\large explore}\xspace}
\newcommand{\ouralgodistance}{\ensuremath{\text{\ouralgo-}\scriptstyle\mathit{D}}\xspace}
\newcommand{\ouralgocover}{\ensuremath{\text{\ouralgo-}\scriptstyle\mathit{C}}\xspace}

\newcommand{\expecteddivdist}{\ensuremath{\bar{D}}}
\newcommand{\expecteddivcov}{\ensuremath{\bar{C}}}

\title{Relevance meets Diversity: A User-Centric Framework for Knowledge Exploration through Recommendations}

\author{Erica Coppolillo}
\orcid{0000-0002-4670-8157}
\affiliation{%
  \institution{University of Calabria}
  \department{Department of Computer Science}
    \city{Rende}
    \country{Italy}
}
\additionalaffiliation{%
  \institution{ICAR-CNR}
    \city{Rende}
    \country{Italy}
}
\email{erica.coppolillo@unical.it}

\author{Giuseppe Manco}
\orcid{0000-0001-9672-3833}
\affiliation{%
  \institution{ICAR-CNR}
  \city{Rende}
  \country{Italy}
}
\email{giuseppe.manco@icar.cnr.it}

\author{Aristides Gionis}
\orcid{0000-0002-5211-112X}
\affiliation{%
  \institution{KTH Royal Institute of Technology}
  \department{Division of Theoretical Computer Science}
  \city{Stockholm}
  \country{Sweden}
}
\email{argioni@kth.se}

\begin{document}

\begin{abstract}
Providing recommendations that are both \emph{relevant} and \emph{diverse} 
is a key consideration of modern recommender systems.
Optimizing both of these measures presents a fundamental trade-off, 
as higher diversity typically comes at the cost of relevance, 
resulting in lower user engagement. 
Existing recommendation algorithms try to resolve this trade-off
by combining the two measures, relevance and diversity, into one aim
and then seeking recommendations that optimize the combined objective, 
for a given number of items to recommend. 
Traditional approaches, however, do not consider
the user interaction with the recommended items.

In this paper, we put the \emph{user} at the central stage, 
and build on the interplay between \emph{relevance}, \emph{diversity}, and \emph{user behavior}.
In contrast to applications where the goal is solely to maximize engagement, 
we focus on scenarios aiming at maximizing the total amount of knowledge encountered by the user. 
We use diversity as a surrogate of the amount of knowledge obtained by the user while interacting with the system, 
and we seek to maximize diversity. 
We propose a probabilistic user-behavior model in which users keep interacting with the recommender system
as long as they receive relevant recommendations, 
but they may stop if the relevance of the recommended items drops.
Thus, for a recommender system to achieve a high-diversity measure, 
it will need to produce recommendations that are \emph{both relevant and~diverse}. 

Finally, we propose a novel recommendation strategy 
that combines relevance and diversity by a copula function. 
We conduct an extensive evaluation of the proposed methodology over multiple datasets, 
and we show that our strategy outperforms several state-of-the-art competitors.
Our implementation is publicly available~\footnote{\url{https://github.com/EricaCoppolillo/EXPLORE}}.
\end{abstract}

\maketitle

% FIXME
%\input{movielens-1m/clayton_alphas_table}
%\input{movielens-1m/ali_alphas_table}

\begin{figure}[t]
\includegraphics[width=\columnwidth]{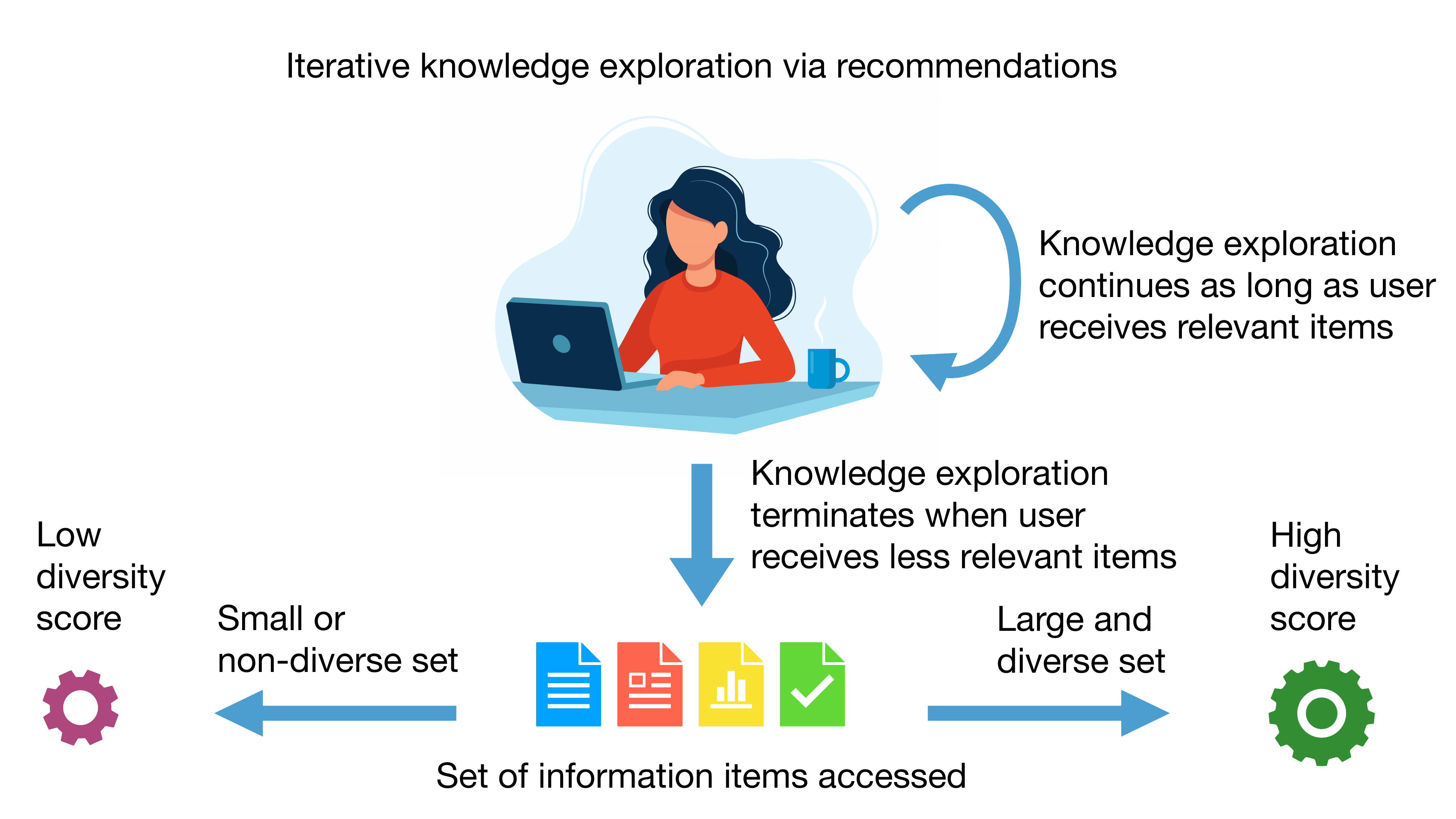}
\caption{The knowledge-exploration process, illustrating the interplay 
among \textit{relevance}, \textit{diversity}, and \textit{user behavior}.}
\label{fig:toy_example}
\end{figure}
\section{Introduction}
\label{sec:introduction}

%Recommender systems are widely used for discovering new information and broadening our knowledge, 
Recommender systems play a significant role in helping users discover new information and expand their knowledge base. 
Notable examples are the adoptions of recommendations for 
%for instance, 
finding news articles or books to read~\cite{alharthi2018survey,wu2023personalized,zihayat2019utility}, 
listening to enjoyable music~\cite{hansen2020contextual,schedl2015music}, 
visiting interesting locations~\cite{xie2016learning,yin2013lcars},
and more.
Recommender systems aim to predict and leverage users' interests to identify 
the portions of the catalog that match them, thus enabling efficient exploration 
of vast volumes of information and offering benefits 
ranging from increased personalization and user satisfaction to improved engagement 
and resource efficiency. 
%They are crucial especially in scenarios where the volume of information available is vast, creating a paradox where effective knowledge exploration is only achievable through intelligent filtering.

Recommenders are primarily focused on maximizing {relevance}. However, from the standpoint of knowledge exploration, incorporating {diversity} into recommendations adds significant value, as emphasized in earlier research~\cite{evaluating_CF, Smyth2001SimilarityVD}.
%Despite initially focusing on predicting users' interests and maximizing {relevance}, 
%{diversity} brings significant added value for recommendations, 
%as pointed out early on~\cite{evaluating_CF, Smyth2001SimilarityVD}.
%\erica{This awareness grew progressively, yielding to several improvements in the last decade~\cite{on_unexpectedness, improving_aggregate, evaluating_novel_recs, diversity_top_n_recs, rank_relevance}. Also, several online and targeted user studies assessed the increase in user satisfaction when diversity is incorporated into the list of suggested items~\cite{diversity_top_n_recs, Castells2022}. For example, Allison et al.~\cite{algorithmic_confounding} show that, if certain objectives (including diversity) are not taken into account, the interactions between users and recommender systems are prone to homogenization and, consequently, low utility.}
Indeed, providing diverse recommendations can be critical in mitigating detrimental consequences,
such as being trapped in \emph{rabbit holes}
% \footnote{\erica{With the term \textit{rabbit holes}, we do not refer necessarily to the phenomenon of radicalization in a political context, but to a broader concept, in which the algorithm leads the user to consume limited types of content. 
%On the other hand, radicalization can be a topic related to diversity as well, since users could become exposed to partial information within a narrow community, which leads their beliefs and experiences to the extremes.}} 
in platforms like Youtube~\cite{rabbit_hole, the_great_radicalizer, ribeiro2020auditing, Haroon2022YouTubeTG} 
or Reddit~\cite{phadke2022pathways}, 
where the algorithm may lead the user to consume limited types of content.
% where users can fall into conspiracy discussions~\cite{phadke2022pathways}
% or promote the spread of misinformation~\cite{echo_chamber_fake_news, Trnberg2018EchoCA, spreading}.
% or follow polarized communities~\cite{Jiang2021SocialMP}.

% To balance \emph{relevance} and \emph{diversity}, 
% existing methods combine the two measures into one objective, 
% and proceed to optimize the combined objective.
% Such approaches, however, do not take into account the \emph{user behavior} 
% and their interaction with the list of recommended items.
% For instance, a fixed number of interactions is generally assumed between the user and the algorithm, 
% and the reactions of the user during such interactions is typically disregarded. In this respect, a user could refuse the recommended items and abandon the exploration process.
% %For instance, the combined objective is optimized for a fixed number of recommended items, while the user may end up interacting with a completely different number of~items.
To achieve a balance between relevance and diversity, current methods merge these two metrics into a single objective for optimization. However, they overlook user behavior and how users interact with the recommended list of items. For instance, typical approaches assume a fixed number of interactions between the user and the algorithm, disregarding any reactions or refusals from the user during the exploration process. Indeed, users might reject recommended items and quit the process.

%\erica{Add a "warm-up" about the use of simulation instead of users study}

In this paper we propose a new framework for recommender systems,
where we place the user at the forefront. 
%Compared to methods that solely seek to maximize user engangement, the proposed framework is tailored to maximize the total amount of knowledge obtained by the user while he/she examines the list of recommendations. 
%\erica{Try rephrasing to give the idea of a "definition" of knowledge exploration. Maybe we can also refer to Section 3 (subsection "user model"), where we formally define the exploration process.} 
%In our framework, w
%
We consider the interaction of the user with the algorithm
to be a \emph{knowledge-exploration task}, where recommendations enable exploration. 
The interaction of the user with the system is guided via a \emph{user-behavior model}, 
i.e., the propensity of a user to accept or reject recommendations according to their preferences and patience.
As the objective is to maximize the amount of knowledge that 
a user acquires during exploration,  
we model the knowledge accrued by the user using a \emph{diversity} measure, 
which we consequently aim at maximizing. 
Notably, although diversity is the sole optimization objective, 
the coupling of the exploration task with the user-behavior model 
implies that the recommendation system is required to produce recommendations that are \emph{both relevant and~diverse}. 
%\erica{Maybe we can rephrase to avoid repeating "diversity".}

We illustrate the proposed concept of \emph{``knowledge exploration via recommendations''} with the following example.

\begin{figure}
\centering
\begin{subfigure}[b]{0.3\columnwidth}
        \centering
        \includegraphics[width=0.25\columnwidth]{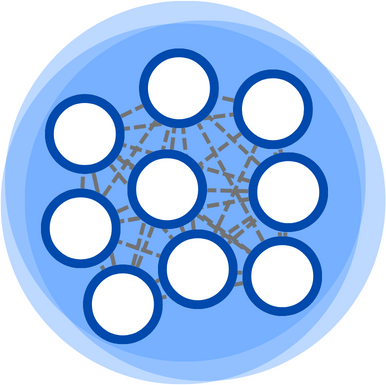}
    \caption{}
    \end{subfigure}
    \begin{subfigure}[b]{0.32\columnwidth}
        \centering
        \includegraphics[width=\columnwidth]{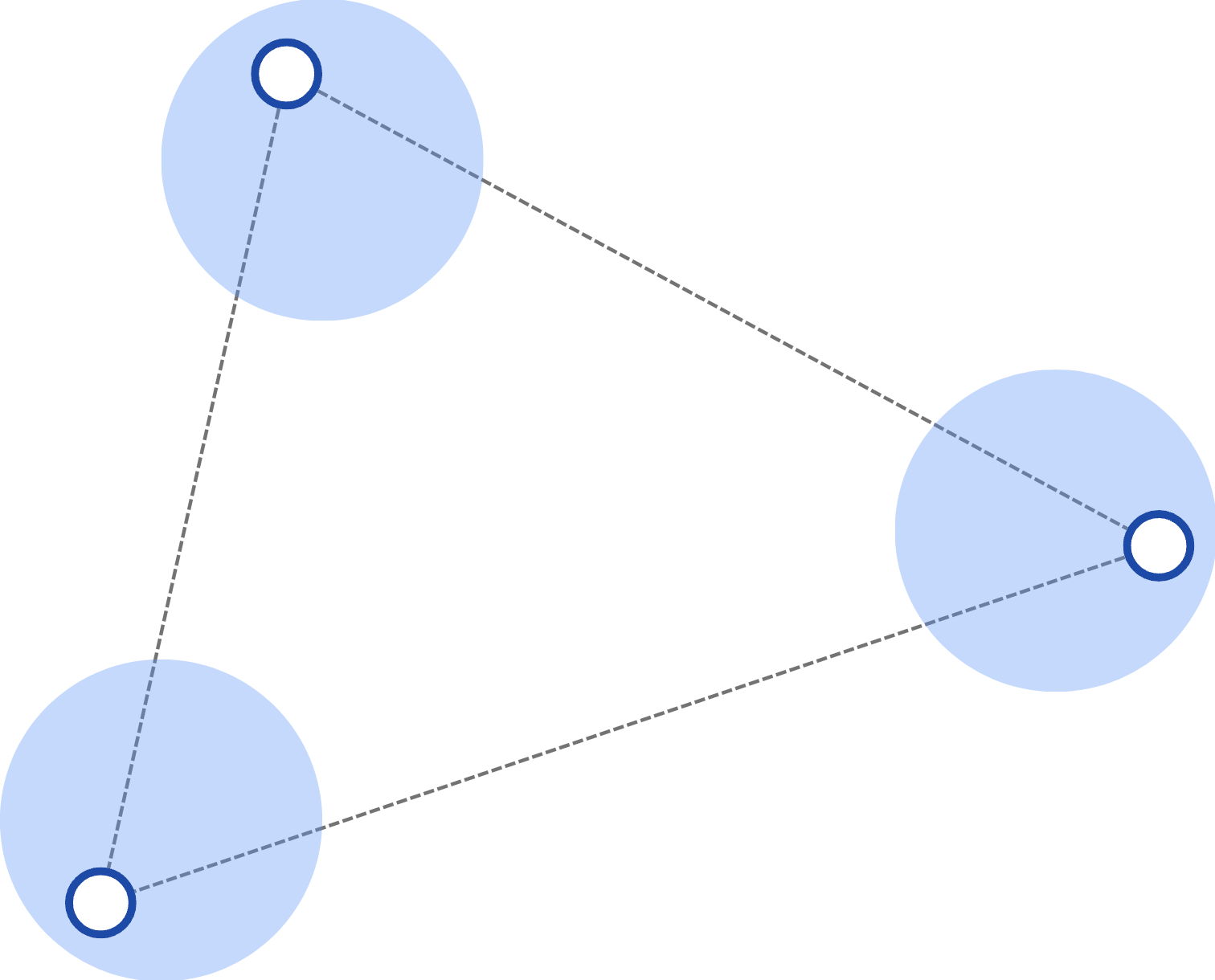}
    \caption{}
    \end{subfigure}
    \begin{subfigure}[b]{0.32\columnwidth}
        \centering
        \includegraphics[width=0.7\columnwidth]{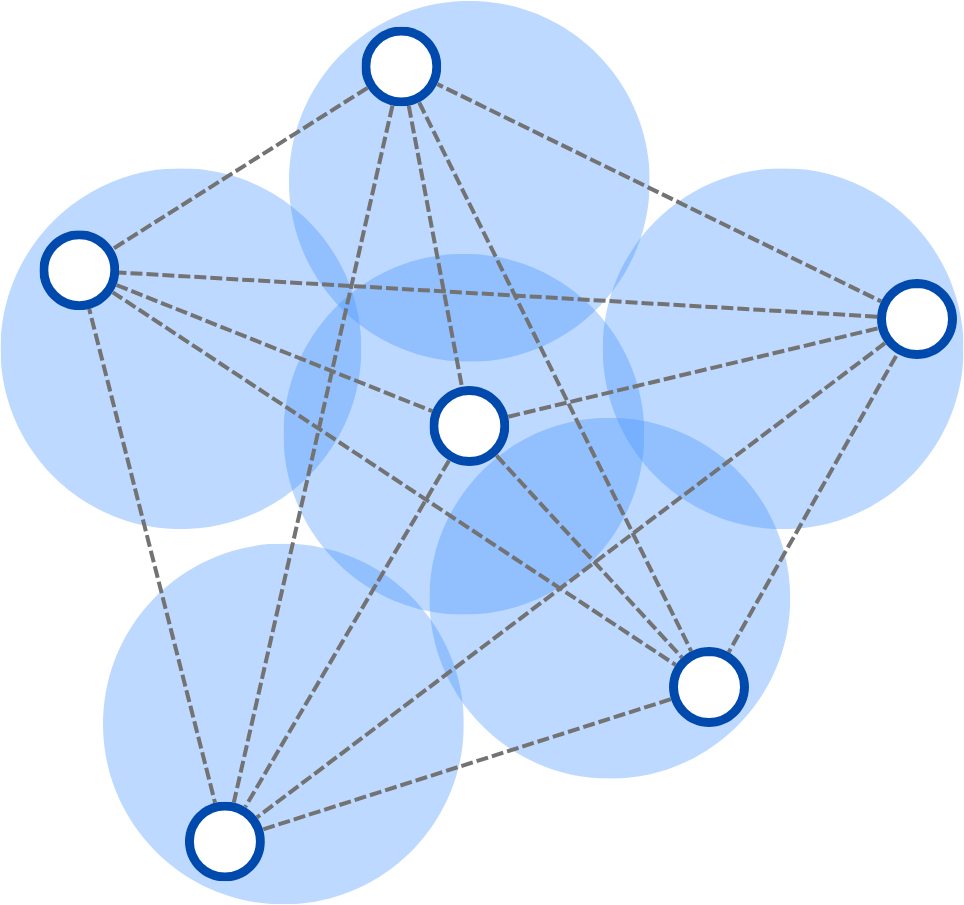}
    \caption{}
    \end{subfigure}
\caption{
Illustration of the impact of different recommendation strategies. 
White points are recommended items, blue circles indicate information coverage.
(a) High relevance, low diversity (e.g., all about `technology');
(b) High diversity, likely non-relevant (e.g., `technology,' `religion,' `lifestyle'); 
(c) Optimal balance: Relevant and diverse, keeping user engaged
(e.g., `technology,' `science,' `engineering.').
}
% \caption{Illustration of the effect of different recommendation strategies.
% The white points represent recommended items.
% The blue circle represent the ``information content'' of recommended items.
% \giuseppe{We should clarify what "information content" means cause it's a bit ambiguous. Maybe we can use "Information coverage", to denote the span that it covers (and hence the possible overlaps that allow propagation of relevance). Possible rewriting:  The blue circle represent the ``information coverage'' of recommended items like, e.g., the topics they are related to. }
% The underlying space (here 2-$d$) represents semantic distances between recommended items.
% (a) Highly relevant recommendations can keep the user engaged but
% the recommendations are likely too similar to each other, and the overall diversity is low\giuseppe{, e.g. cause they are all about "technology"}.
% (b) Highly diverse recommendations \giuseppe{(e.g., covering technology, arts and lifestyle)} are likely to be non relevant,
% and the user will loose interest after a small number of recommendations.
% (c) A desirable trade-off between relevance and diversity; 
% the recommendations are sufficiently relevant and diverse for the user to stay engaged 
% and consume a diverse information~diet.
% \aris{these are all great suggestions! Let's implement them.}
% }
\label{fig:diversity_scenarios}
\end{figure}

%\iffalse
\spara{Example.}
Alice interacts with a news recommender system for finding 
interesting news articles to read.
The knowledge-exploration process is iterative, and is depicted in Figure~\ref{fig:toy_example}. 
At each step, the system recommends a set of news articles to Alice, 
and Alice clicks on some article to read. 
At some point, Alice can decide to quit, 
either because she received enough information, 
or because the recommendations are not very interesting to her, 
or simply because she got bored.
Our goal is to design a recommender system that maximizes the amount of knowledge received by Alice. 
%While common approaches focus on maximizing engagement and time spent, we propose prioritizing the total amount of information received, modeled using diversity.
The challenge is to strike a balance between diversity and relevance to keep Alice engaged 
while exploring interesting topics, 
avoiding scenarios where recommendations are either too focused (Figure~\ref{fig:diversity_scenarios}(a)) 
or too diverse and irrelevant (Figure~\ref{fig:diversity_scenarios}(b)). 
Our aim is to create an ideal scenario (Figure~\ref{fig:diversity_scenarios}(c)) 
where Alice explores many relevant yet diverse topics, enriching her~knowledge.

% Moreover, the lack of diversity in information can lead to detrimental consequences by promoting the spread of misinformation~\cite{echo_chamber_fake_news, Trnberg2018EchoCA, spreading}: on Twitter and Facebook, the polarized communities were proven to be the main source of fake news proliferation related to the COVID-19 pandemic~\cite{Jiang2021SocialMP}, the 2016 US presidential election~\cite{us_election}, and the recent Russia-Ukrain war~\cite{russia_ukrain_war}.  

% \smallskip
% Motivated by the previous example, \erica{Adding research questions?
% \begin{itemize}
%     \item \textbf{RQ1}: How can we model the interplay among diversity, relevance, and user behavior? 
%     \item \textbf{RQ2}: Can we devise a strategy for delivering accurate yet diverse recommendations, given a baseline black-box recommender system? 
%     \item \textbf{RQ3}: How does the user behavior affect the interactions within the recommender system?
% \end{itemize}
% }
% Motivated by the previous example, we propose a novel framework to model the behavior of the user
% while exploring information via recommendations. 
% Relevance is taken into account to model termination of exploration, 
% but the overall quality is measured by diversity.
% We instantiate our model using two standard notions of diversity, 
% one based on coverage and the other based on pair-wise distances~\cite{dum, mmr, dpp}.

Motivated by the previous example, we propose a novel framework where relevance governs the termination of exploration, while the overall quality is measured by diversity.
We instantiate our model using two standard notions of diversity, 
one based on coverage and the other based on pair-wise distances~\cite{dum, mmr, dpp}.
Both diversity notions, coverage and pairwise distances, 
can be defined using an underlying space of user-to-item ratings or categories/topics. 
%Other diversity notions can also be incorporated into our framework, e.g., pair-wise distances of item embeddings obtained by representation learning, but we leave this for further study, as it is not our focus.

Finally, we propose a novel recommendation strategy 
that combines relevance and diversity by a copula function. 
We perform an extensive evaluation of the proposed framework and strategy using five benchmark datasets publicly available, and show that our strategy outperforms several state-of-the-art competitors.

% This given, in this work, we address the presented problem by ($i$) developing a novel simulation framework that relies on both accuracy and diversity, ($ii$) defining two measures to be applied to the final set of user interactions, and ($iii$) implementing two recommendation strategies that prove to be effective in this context.

Our contributions are summarized as follows:
\begin{itemize}
    \item We develop a user-centric model for knowledge exploration via recommendations; 
    our framework takes into consideration the interplay among relevance, diversity, and user behavior.
    \item We instantiate our model with two diversity measures, defined over user-to-item ratings or categories/topics.
    \item We propose a recommendation strategy that accounts for both diversity and relevance when providing suggestions.
    \item We conduct an extensive analysis over multiple benchmark datasets and several competitors to show the effectiveness of our proposal in the suggested framework.
\end{itemize}

The rest of the paper is structured as follows. Section~\ref{sec:related} presents the related work in terms of user modeling and diversity in recommendations. Section~\ref{sec:problem} presents our problem definition and methodology. In Section~\ref{sec:our_strategies} we present our recommendation strategy. Experimental results are reported in Section~\ref{sec:experiments}, and finally Section~\ref{sec:conclusion} concludes the paper and provides pointers for future extensions. 

\section{Related work}
\label{sec:related}

% \erica{as reported by Castells et al.~\cite{Castells2022},  the importance of diversity has been early recognized~\cite{evaluating_CF, Smyth2001SimilarityVD} as an added value for recommendations~\cite{first_diversification}. This awareness grew progressively, yielding to several improvements in the last decade~\cite{on_unexpectedness, improving_aggregate, evaluating_novel_recs, diversity_top_n_recs, rank_relevance}. Also, several online and targeted user studies assessed the increase in user satisfaction when diversity is incorporated into the list of suggested items~\cite{diversity_top_n_recs, Castells2022}. For example, Allison et al.~\cite{algorithmic_confounding} show that, if certain objectives (including diversity) are not taken into account, the interactions between users and recommender systems are prone to homogenization and, consequently, low utility.}

\spara{User modeling in recommender systems.} 
The effects of user behavior in recommender systems, in terms of novelty and diversity, have gained a lot of attention in recent years. Analysis can be conducted by either running user studies~\citep{Lee2014ImpactOR, recsys_sales_diversity, sales_diversity_impact, Zhu2018TheIO}, or by means of simulation~\citep{simulating_impact, diversity_under_users_modeling, Yao2021MeasuringRS}. %Examining the decisions made by real users can lead to more reliable results, but it introduces the challenge of developing a functional recommendation system and conducting extensive user~studies. 
Analyzing the choices made by actual users can yield more dependable outcomes; however, it also requires creating an effective recommendation system and engaging users for conducting comprehensive studies. 
%which can be highly consuming, both in terms of time and resources. 

On the other hand, simulating user choices is a more straightforward method, 
allowing for testing several system configurations at no expense. 
However, it requires a realistic model of \emph{user behavior}.
To address this challenge, several user-behavior models have been proposed in the literature. 
\citet{HAZRATI2022102766} model the probability that a user picks a recommended item 
as being proportional to the utility of the item.
%  scaled by a factor representing the uncertainty in its estimation and boosted by a multiplicative factor.
Similarly, \citet{siren} propose a simulation framework in which users decide to interact with a certain number of items per iteration, according to their given preferences. 
\citet{diversity_under_users_modeling} present three different user-behavior models, 
by imposing that users either blindly follow recommendations and choose the most popular items, 
or completely ignore suggestions and pick items randomly. 

The aforementioned models present certain limitations, 
namely users necessarily have to pick an item, i.e., they cannot leave the application, 
and second, the selection probability stays constant over time.
% i.e., attrition of interest with non-relevance recommendations, or with time is not modeled.
We overcome these limitations by modeling a \textit{quitting probability}, 
according to which users can interrupt their interaction with the recommender system. 
We assume that the quitting probability depends on the utility of the recommended items
and on the user \textit{patience}, which degrades over time. 

Notably, with our framework, we leverage the intrinsic interplay among relevance, diversity, and user behavior, 
since successful recommendation strategies need to ensure that they provide recommendations that are both relevant and diverse.

\para{Diversity in recommendation.} 
Diversity in recommendations has been acknowledged as a crucial issue \cite{Castells2022, evaluating_CF, Smyth2001SimilarityVD, first_diversification}, and over the past decade, it has received considerable attention \cite{on_unexpectedness, improving_aggregate, evaluating_novel_recs, diversity_top_n_recs, rank_relevance}.
Several online and targeted user studies assessed the increase in user satisfaction when diversity is incorporated into the list of suggested items~\cite{diversity_top_n_recs, Castells2022}. For example, Allison et al.~\cite{algorithmic_confounding} show that, if diversity (besides other objectives) is not taken into account, the interactions between users and recommender systems are prone to homogenization and, consequently, low utility.

The challenge of striking a balance between diversity and relevance has been explored both in the context of recommender systems and in the broader domain of information retrieval.
For instance, one of the most popular methods in the literature of information retrieval is the \emph{maximal marginal relevance} (MMR)~\cite{mmr}. It employs a weighted linear combination of scores that evaluate both utility and diversity, offering a systematic way to address this critical aspect. 
In the specific context of recommender systems, \citet{first_diversification} introduced one of the earliest methods for enhancing diversity. They use a greedy selection approach, where they pick items that minimize the similarity within a recommended list.
\citet{accuracy-diversity_dilemma} present a solution based on random walks for the so-called \textit{accuracy-diversity dilemma}, i.e., the challenge in finding a profitable trade-off between the two measures. 
This concept is also known as \textit{calibration}, as mentioned by~\citet{calibration}, and refers to the algorithm's capability to produce suggestions that do not under-represent (or ignore) the user's secondary areas of interest. 

Several re-ranking strategies have also been introduced:
\citet{dum} propose to greedily select items by maximizing the utility of a submodular function; 
\citet{dpmf} suggest to optimize the diversity loss of items using probabilistic matrix factorization; 
\citet{dpp} propose a determinantal point process (DPP) to re-rank the recommended items so as to maximize the determinant on the items' similarity~matrix. \citet{diverse_content} investigate the impact of diversity on music consumption, and propose two innovative models: a feed-forward neural ranker that produces dynamic user embedding, and a reinforcement learning-based ranker optimized on the track relevance. 
%The former incorporates the user’s previous sessions to produce a dynamic user embedding; while the latter defines the reward based on whether the user found the track to be relevant.
\emph{Reinforcement learning} is indeed a suitable solution for addressing the diversity problem. It plays a role in the work by \citet{multi-armed}, 
where diversity is induced by adopting multi-armed bandits in the elicitation phase; and in the online learning framework proposed by \citet{linear_bandits}, where diversification is obtained by carefully balancing the exploration and exploitation of users' preferences and interests.
Notably, these reinforcement learning-based approaches typically require a lengthy training phase, which can often be prone to stability issues. 
% This is in contrast to our approach, which may offer a more efficient and stable solution.

Several other \emph{neural-network models} have been applied to address the diversity problem. 
\citet{diversification_vae_based} adopt a \emph{variational auto\-encoder} to induce targeted (i.e., topical) diversity. 
\citet{enhancing_adaptivity} propose a \emph{bilateral branch network} to achieve a good trade-off between relevance and diversity, defined at either domain or user level. 
\citet{dgcn} present a \emph{graph neural network} for diversified recommendations, 
where node neighbors are selected based on inverse category frequency, 
together with negative sampling for inducing diverse items in the embedding space.

% \aris{How do these NN methods compare with our method?
% Do they compete with our method?}
% \erica{They do, since they optimize diversity (either in terms of distance or coverage) in their loss function. Indeed, we use the last cited one as a competitor.}

% We argue that the reported approaches are either highly demanding in terms of time and/or computational resources, or rely on some heuristics not theoretically justified.

In contrast to most of the approaches mentioned earlier, the recommendation strategy we introduce, \ouralgo, does not necessitate any form of training or hyperparameter tuning, it is computationally efficient, and is shown to provide both highly relevant and diverse suggestions.

\section{User model and problem formulation}
\label{sec:problem}

%\aris{Need to make sure that in the rest of the paper ``information exploration'' is replaced by ``knowledge exploration''.}

{
%\scriptsize
\begin{algorithm}
\caption{Simulation process for user \auser}
\label{alg:simulation}
\hspace*{-5.7cm}\textbf{Input}: \auser, \items, \strategy, \relscore \\
\hspace*{-6.62cm}\textbf{Output}: \itemset
\begin{algorithmic}[1]
\small
\State $\itemset \gets \emptyset$
\State $quit \gets$ False
\While{\textbf{not} $quit$}

\State $\ilistk = [i_i, i_2, ..., i_\ilistsize] \gets \strategy(\relscore(\auser, \items \setminus \itemset), \itemset)$
\State \textit{examining} \ilistk $\gets$ Algorithm~\ref{alg:user_behavior}
\If{\auser does not quit}
    \State $\anitem \gets$ picked item
    \State $\itemset \gets \itemset \cup \{\anitem\}$
\Else
    \State $quit \gets$ True
\EndIf
\EndWhile
\end{algorithmic}
\end{algorithm}
}

{
%\scriptsize
\begin{algorithm}
\caption{User behavior at step $t$}
\label{alg:user_behavior}
\hspace*{-6.9cm}\textbf{Input}: \ilistk \\
\hspace*{-5.2cm}\textbf{Output}: $\anitem \in \ilistk$ or \textit{quits}

\begin{algorithmic}[1]
\small
\State \textit{interest} $\gets$ False
\For{$j=1, \ldots, \ilistsize$}
    \State $\anitem \gets \ilistk[j]$
    % \State \textbf{quits} $\gets$ with probability (1 - $\theta_j$)
    \State \textit{quitting} $\gets$ with probability \quitprobk
    \If{\auser quits}
        \State \textbf{return}
    \Else
        \State \textit{examining} \anitem $\gets$ with probability \bufferprobi
        \If{\anitem is interesting}
            \State \textit{interest} $\gets$ True
        \EndIf
    \EndIf
\EndFor
\If{not \textit{interest}}
    \State \textbf{return}
\EndIf

\For{$j=1, \ldots, \ilistsize$}
    \State $\anitem \gets \ilistk[j]$
    \State \textit{consuming} $\anitem \gets$ with probability \selectprobi
    \If{\auser consumes \anitem}
        \State \textbf{return} \anitem %\textbf{stops}
    \EndIf
\EndFor
\end{algorithmic}
\end{algorithm}
}

We consider a typical recommendation setting in which we have a set of \nousers users \users 
and a set of \noitems items \items. 
We also consider a function $\relscore:\users\times\items\rightarrow\reals$ 
that provides us with a relevance score
$\relscore(\auser,\anitem)$, for each user $\auser\in\users$ and item $\anitem\in\items$.
We assume that the function \relscore can be computed by a black-box method, 
and state-of-the-art relevance-scoring functions can be employed, such as content similarity~\cite{Pazzani2007}, collaborative filtering~\cite{Schafer2007}, or a combination of both~\cite{Burke2007}.
Our goal in this paper is to create lists of diverse recommendations using 
as a black box such relevance-scoring functions, rather than devising a new relevance-scoring function~\relscore.

\spara{Item-to-item distance function.}
We next discuss how to define a distance function between pairs of items in \items, which will be used in one of our two diversity definitions. 

Given an item $\anitem\in\items$, we denote by \vectxi the vector of \textit{users} with
\begin{equation*}
    \vectxiu =
    \begin{cases}
      1, & \text{if user \auser interacted with item \anitem,} \\
      0, & \text{otherwise}.
    \end{cases}
\end{equation*}
The vectors $\{\vectxi\}$ can be retrieved by user-log data.
% if available.
A more fine-grained representation of vectors $\{\vectxi\}$ beyond binary is also possible,
for instance, using numerical values that represent the \emph{rating} of user \auser for item \anitem, 
if such information is~available.

An alternative approach is to use categories 
(or keywords, or genres, depending on the application).
In particular, we consider a set of categories \categories, 
and we define \vectyi to be a \textit{category} vector, for item $\anitem\in\items$,  where
\begin{equation*}
    \vectyic =
    \begin{cases}
      1, & \text{if category \acategory relates to item \anitem,} \\
      0, & \text{otherwise}.
    \end{cases}
\end{equation*}

% and $\textbf{y}_i$ be the \textit{categories} vector such that:
% \begin{equation}
%     y_{ic} =
%     \begin{cases}
%       1, & \text{if category $c$ relates to item $i$} \\
%       0, & \text{otherwise}
%     \end{cases}
% \end{equation}

Given two items $\anitem,\anotheritem\in\items$, 
we hence define their \emph{distance} as the \emph{weighted Jaccard distance}
\begin{equation}
    \distance(\anitem, \anotheritem) = 
        1 - \frac{\sum_{w \in \setw} \min\{z_{iw}, z_{jw}\}}{\sum_{w \in \setw} \max\{z_{iw}, z_{jw}\}},
\label{equation:distance}
\end{equation}
where \setw is either the set of users \users or the set of categories \categories, 
and accordingly, 
\vectzi is the \textit{user vector} or the \textit{category vector} of item~\anitem.

Finally, we note that other state-of-the-art distance functions can also be used, 
such as Euclidean distance, cosine similarity, or Minkowski distance~\cite{distance_functions}.
We do not investigate what is the best distance function to be used, 
as this is orthogonal to our study and beyond the scope of this paper.

\spara{Diversity.}
Given a set of items $\itemset\subseteq\items$, we define the 
\emph{diversity} of the set \itemset.
We explore two different definitions of diversity. 

Our first definition is based on the concept of \emph{coverage}. 
It assesses the degree to which the items within \itemset adequately represent the entire range of categories \categories. 
In particular, for a set of items $\itemset\subseteq\items$,  
we define its \emph{coverage-based diversity} as
\begin{equation}
\divcov(\itemset) = 
    \frac{1}{|\categories|}
    \left\| \bigvee\nolimits_{\anitem\in\itemset} \,\vectyi \right\|_0,
\label{equation:coverage-based-diversity}
\end{equation}
where $\|\cdot\|_0$ returns the number of non-zero entries of the binary vector 
$\bigvee_{\anitem\in\itemset} \,\vectyi$.
Notice that the metric \divcov is scaled to fall within the range of 0 to 1, considering the total number of categories in \categories. 
%
% It is worth highlighting that it is influenced by the size of \itemset. In simpler terms, the diversity measure \divcov tends to favor larger \itemset sizes, as they typically cover a wider range of categories. Additionally, \divcov naturally prefers items with extensive coverage.
It is worth highlighting that \divcov favours larger \itemset sizes, as they typically cover a wider range of categories. Additionally, \divcov naturally prefers items that individually provide extensive coverage.

Our second measure of diversity employs the distance function \distance
that we defined in the previous paragraph. 
In particular, for a set of items $\itemset\subseteq\items$ with $|\itemset|\ge 2$, 
we define its \emph{distance-based diversity}~as
\begin{equation}
\divdist(\itemset) = 
    \frac{1}{|\itemset|-1}
    \sum_{\anitem\in\itemset}\sum_{\anotheritem\in\itemset} \distance(\anitem,\anotheritem),
\label{equation:distance-based-diversity}
\end{equation}
and we define $\divdist(\itemset) = 0$, if $|\itemset|< 2$.
Notice that the number of terms in \divdist is quadratic with respect to $|\itemset|$.
By normalizing with $(|\itemset|-1)$ the dependence becomes linear in $|\itemset|$.
As with \divcov, the \divdist metric favors larger sets, 
in addition to favoring items whose distance is large to each other.

\spara{User model.}
A central aspect of our approach is that we aim to evaluate the quality 
of a recommendation algorithm \strategy in the context of the user response to items recommended by \strategy.
We view the user-algorithm interaction as a dynamic 
 knowledge-exploration process, 
in which the algorithm recommends items to the user, 
and the user interacts with the recommended items. 
The knowledge-exploration process continues as long as the recommended items are of interest to the user. 
If the recommended items are not interesting enough 
(meaning, if they have low relevance for the user)
the user may (stochastic\-al\-ly) decide to quit.

To formalize the exploration process between the user and the recommendation algorithm \strategy, 
which is needed to evaluate the quality of \strategy, 
we propose a \emph{user model}.
Our model is specified in terms of a relevance-scoring function \relscore, 
which guides the behavior of the user, 
and in terms of a recommendation algorithm \strategy, 
which enacts the choices within \strategy. % \erica{? of the user?}. 

Our user model, which formalizes knowledge-exploration as an iterative process, is described as follows.

\begin{comment}
\begin{enumerate}
\item 
The set of items that the user interacts with 
during the exploration process is denoted by \seenitems. 
Initially \seenitems is empty.
\item 
In the \iter-th iteration, the recommendation algorithm \strategy forms a list
of items \ilistk, which are presented to the user. 
\item 
With certain probability that depends on the relevance of the recommended items (and which we quantify later),
the user does not find any interesting item in the list \ilistk and quits the exploration process. 
\item 
If the user does not quit, 
with certain probability that depends on the relevance of the recommended items (and which we quantify later),
the user selects an item \anitem from the list \ilistk and interacts with it. 
The item \anitem is added to the set \seenitems.
\item 
Upon quitting, the total score achieved by the recommendation algorithm \strategy is
defined to be $\diversity(\seenitems)$, 
where \diversity is one of our diversity functions, \divcov or \divdist.
\end{enumerate}    
\end{comment}

\begin{enumerate}
\item 
The set of items that the user interacts with 
during the exploration process is denoted by \seenitems. 
Initially \seenitems is empty.
\item 
In the \iter-th step, the recommendation algorithm \strategy generates a list
of items \ilistk to present to the user. The user examines these items in a specified order.
\item 
At any point in the current step, the user has the option to quit. The likelihood of quitting (to be quantified later) depends on two factors: the relevance of the recommended items and the user's patience. If the user fails to find interesting items in list \ilistk or if they stochastically run out of patience, they may opt to conclude the exploration process.
\item 
If the user does not quit, 
with a certain probability that depends on the relevance of the recommended items (and which we quantify later),
they select an item \anitem from the list \ilistk and interact with it. 
The item \anitem is added to the set \seenitems and the exploration process continues. 
\item 
Upon quitting, the total score achieved by the recommendation algorithm \strategy is determined to be $\diversity(\seenitems)$, where \diversity is one of our diversity functions, \divcov or \divdist. This score reflects the diversity in the items the user has interacted with throughout the exploration process. We denote the final number of steps performed by the user as \steps.
\end{enumerate}

Algorithm~\ref{alg:simulation} depicts the overall exploration process.

To fully specify the user model we need to describe in more detail the probability 
that the user selects an item to interact with, 
as well as the probability of quitting the exploration.
Before presenting more details about these aspects of the model, 
we first formalize the problem of designing a recommendation algorithm
in the context of our user model.

\spara{The recommendation task (problem statement).}
The algorithmic problem that we address in this paper is the following.

\begin{problem}
Given a set of items \items, a set of users \users,  
a relevance-scoring function $\relscore:\users\times\items\rightarrow\reals$, 
a diversity function  $\diversity:2^\items\rightarrow\reals$, 
and a user model for knowledge-exploration as the one described in the previous paragraph, 
the goal is to design a recommendation algorithm \strategy
that maximizes the diversity score $\diversity(\seenitems)$
for the set of items \seenitems that a user $\auser\in\users$ interacts with.
\label{problem:strategy-design}
\end{problem}

\spara{Item selection.}
We now discuss step (4) of the iterative knowledge-exploration user model presented in the previous paragraph, 
that is, 
we specify how we model the probability that a user selects an item~\anitem from the list \ilistk to interact with. 
We first assume that a user does not quit exploration, i.e., that they have enough patience to explore the whole \ilistk and that they find at least a relevant item within it
%We first compute the probability that a user quits exploration
(see next paragraph). 
%If a user does not quit, then 
In that case, the user 
selects an item~\anitem from \ilistk with probability proportional to the relevance of \anitem
for that user~\auser, that is, 
$\selectprobi = \frac{\relscore(\auser,\anitem) }{\sum_{\anotheritem \in \ilistk}\relscore(\auser,\anotheritem)}$.
As noted before, 
the selected item~\anitem is added to the set of interacted items \seenitems.

\iffalse
We also assume that \auser cannot be suggested an item that has already interacted with, 
i.e., $\ilistk \cap \seenitems = \emptyset$.

\aris{Do we really need this assumption? I do not see why.}
\fi 

\spara{Quitting exploration.}
Last, we discuss step (3) in our user model, 
that is, how we model the probability that a user quits the exploration process.
A sensible model for the quitting probability is crucial in our knowledge-exploration model, 
since we want to mimic user behavior as realistically as possible. 
In particular, we take into consideration two aspects: 
(${i}$) users decide to interact with the recommended items according to their relevance; and 
(${ii}$) users' desire for exploration degrades with time, i.e., users get bored.

In the model we propose, a user examines the items in the list~\ilistk sequentially.
Upon examining an item $\anitem\in\ilistk$ the user decides with probability \quitprobk to quit exploration due to worn out at step \iter. We refer to this as the \textit{weariness} probability. 
The weariness probability \quitprobk, which is discussed in more detail below, 
models the user's decline of interest in exploration as a function of time, and it
depends on the current step \iter in the exploration process.

If the user does not quit, 
they decide whether item \anitem is interesting to explore. 
The latter is decided again stochastic\-ally with Bernoullian probability \bufferprobi, which is a function of the relevance score $\relscore(\auser,\anitem)$.\footnote{In our experiments, \bufferprobi is obtained by normalizing $\relscore(\auser,\anitem)$ into the $[0, 1]$ interval by considering the maximum relevance range.}
Thus, the probability \bufferprobi models the user's interest in an item
according to its relevance.
The examination of the list \ilistk continues until the user decides to quit
or decides that there is at least one item that is interesting to explore.
Thus, the probability that the user quits examining the list \ilistk without identifying any item to explore is
\begin{equation}
\begin{split}
\label{round_quitting_prob}
\quitatlistk = 
    & \ \mbox{\{pr.\ quitting after the first item\}} \ + \ \ldots \  + \\
%     & \ \mbox{\{pr.\ quitting after the second item\}} 
    & \ \mbox{\{pr.\ quitting after the last item\}} \\
  = & \ \sum_{\anotheritem=1}^{|\ilistk|} 
                \quitprobk 
                (1-\quitprobk)^{\anotheritem-1} 
                \prod_{\anitem=1}^{\anotheritem-1} (1-\bufferprobi).
\end{split}
\end{equation}
The last ingredient in our model is to quantify the weariness probability \quitprobk at step \iter. 
This probability models the user's increasing impatience or boredom as their interaction continues. To achieve this, we employ the Weibull distribution~\cite{papoulis2002probability}, which has been previously used to model web page dwell times and session lengths in web page navigation~\cite{liu2010understanding}.

The Weibull distribution is described by two parameters,
$\lambda$ and $\gamma$, 
where $\lambda>0$ is the scale parameter and $\gamma>0$ is the shape parameter of the distribution. 
In particular, we set the weariness probability \quitprobk by resorting to the discrete version of the Weibull Distribution~\cite{weibull_handbook}:
%, which guarantees \quitprobk to be in the range $[0, 1]$:
\begin{equation}
    \quitprobk = 1 - q^{(\iter+1)^\gamma - \iter^\gamma},
\end{equation}
where $q = e^{-1/\lambda^\gamma}$, $0 \leq q \leq 1$.

\noindent

The shape parameter $\gamma$ controls the ``aging'' of the process. 
For $\gamma=1$ the weariness probability remains constant, 
and the resulting distribution becomes an exponential distribution, while 
for $\gamma>1$, the weariness probability increases over time --- 
modeling the tiredness of the user.\footnote{
For $\gamma<1$ the weariness probability decreases over time. 
}

We can use the analytical properties of the Weibull distribution
to obtain the expected number of steps in the exploration process, 
for the case that all recommended items 
are maximally relevant, i.e., $\bufferprobi = 1$ for all $\anitem\in\ilistk$.
In this case, there will be exactly one coin-flip for quitting exploration
for each list \ilistk, and thus, $\quitatlistk = \quitprobk$, for all \iter. 
The overall quitting probability \totalquit is then
\begin{equation}
\begin{split}
    \totalquit = & \ \mbox{\{pr.\ quitting at step 1\}} \ + \ \ldots \  + \\
    & \ \mbox{\{pr.\ quitting at step }\iter\} \ + \ \ldots \ \\
               = & \sum_{\iter=1}^{\infty} \quitatlistk \prod_{j=0}^{\iter-1} (1 - Q_j) \\
               = & \sum_{\iter=1}^{\infty} \left(1 - q^{(\iter+1)^\gamma - \iter^\gamma}\right) \prod_{j=0}^{\iter-1} q^{(j+1)^\gamma - j^\gamma} \\
               = & \sum_{\iter=1}^{\infty} \left(1 - q^{(\iter+1)^\gamma - \iter^\gamma}\right) q^{\iter^\gamma} \\ 
               = & \sum_{\iter=1}^\infty q^{\iter^\gamma} - q^{(\iter+1)^\gamma}.
\end{split}
\end{equation}

The expected number of steps \expectation{\mathrm{steps}}
examined by a user before quitting 
(or equivalently, the number of items in \seenitems) 
is hence given~by
\begin{equation}
    \expectation{\mathrm{steps}} = \sum_{\iter=1}^\infty \iter \Big( q^{\iter^\gamma} - q^{(\iter+1)^\gamma} \Big).
\label{eq:expected-steps}
\end{equation}

Although lacking closed-form analytical expressions, \citet{Khan1989OnEP} show that it is bounded by the expectation $\mu = \lambda\,\Gamma(1 + {1}/{\gamma})$ of the Weibull distribution in the continuous setting~\cite{papoulis2002probability} as
% \begin{equation*}
%     \expectation{\steps} - 1 < \mu < \expectation{\steps}
% \end{equation*}
% %where $\expectationcontinuous{x} = \lambda\,\Gamma(1 + {1}/{\gamma})$ is the expectation of the Weibull distribution in the continuous setting~\cite{papoulis2002probability}. 
% By rearranging the terms, %~\footnote{$\expectation{x} -1 < \expectationcontinuous{x} < \expectation{x} = -1 < \expectationcontinuous{x} - \expectation{x} < 0 = \\ -1 -\expectationcontinuous{x} < -\expectation{x} < -\expectationcontinuous{x} = \expectationcontinuous{x} < \expectation{x} < \expectationcontinuous{x} + 1$.}, 
% we finally obtain:
\begin{equation}
     \mu < \expectation{\mathrm{steps}} < \mu + 1, 
\end{equation}
which provides an algebraic relationship between the $\lambda$ parameter of the Weibull distribution and the admissible range for the expected number of steps.

Note that, if the relevance of the recommended items is less than 1, 
it is possible to get more than one coin-flip for quitting exploration in each list \ilistk. 
In this case, the right-hand side of Equation~(\ref{eq:expected-steps}) provides 
an upper bound on the expected number of steps during exploration.

A notation table can be found in the Appendix (Table~\ref{tab:notation}).

\iffalse
As a result, the probability density for the total number of steps $x$
of the user before quitting is given by 
\[
f(x;\lambda ,\gamma)=
    {\begin{cases}
        {\frac {\gamma}{\lambda }}\left({\frac {x}{\lambda }}\right)^{\gamma-1}e^{-(x/\lambda )^{\gamma}}, & \mbox{if } x\geq 0,\\
        0,&\mbox{if } x<0.\end{cases}}
\]
The expected number of steps before quitting is then
$\expectation{\mathrm{steps}} = \lambda\Gamma(1 + \frac{1}{\gamma})$.
\fi

\spara{Remarks on the proposed model.}
We observe that, as intended, our model captures both the 
relevance of the recommended items 
and the natural tiredness of users with exploration over time.
For fixed values of the Weibull distribution parameters $\lambda$ and $\gamma$, 
which control scaling and aging, 
the users' time for exploration increases with the relevance of the recommended items. 
Furthermore, the \textit{ordering} of the items in the list \ilistk is important,
and thus, we are viewing the recommendation list as a sequence, and not just as a set. 
This aspect would have implications on how to pick the appropriate recommendation strategy, 
but also on the objective (diversity) function, since it can affect the choices of the user.

\section{Recommendation strategy}
\label{sec:our_strategies}

In this section, we present our recommendation strategy
for the proposed knowledge-exploration framework.
Recall that the recommendation task is displayed as Problem~\ref{problem:strategy-design}.

The core of the problem is to construct a list of recommendations \ilistk of size $\|\ilistk\|=\ilistsize$
for the \iter-th step of exploration, for a given user $\auser\in\users$.
We assume that \seenitemsk is the set of items that the user has interacted with at step \iter, 
where $\seenitems_1=\emptyset$.
We define $\candidatesk = \items \setminus\seenitemsk$ to be set of items that are available
for recommendation, that is, all items except the ones that the user has already interacted with.

For a user \auser and each item in the candidate set $\anitem \in \candidatesk$
we consider its relevance score $\relscore_\anitem = \relscore(\auser,\anitem)$
and its \emph{marginal diversity} 
\begin{equation}
\idiversity_\anitem = \diversity(\seenitemsk\cup\{\anitem\}) - \diversity(\seenitemsk), 
\label{equation:marginal-diversity}
\end{equation}
with respect to the interaction set \seenitemsk, where \diversity $\in$ \{\divdist, $\divcov$\}.
We denote $\idiversity_\anitem = \sumdistance_\anitem$ when the distance diversity function \divdist is used, 
and $\idiversity_\anitem = \sumcoverage_\anitem$ when the coverage diversity function \divcov is used.
Intuitively, $\sumdistance_\anitem$ represents the distance of \anitem from all the items in the interaction set 
\seenitemsk, while~$\sumcoverage_\anitem$ represents the additional coverage that~\anitem provides.\footnote{At the beginning of the exploration process (when \seenitemsk = $\emptyset$), if $\idiversity_i = \sumdistance_i$, the strategy samples a highly relevant item $i_r$ so that $\sumdistance_i = \distance(i, i_r)$; if $\idiversity_i = \sumcoverage_i$, then $\sumcoverage_i = \textbf{y}_i$, thus picking the item that individually provides the highest coverage.} Given $\score_\anitem \in \{\relscore_\anitem, \idiversity_\anitem\}$, 
we also denote the min-max normalization of the score \score as 
$\normscore_\anitem = ({\score_\anitem - \score_\mathit{min}})/({\score_\mathit{max} - \score_\mathit{min}})$, 
where $\score_\mathit{max}$ and $\score_\mathit{min}$ are the maximum and minimum values of \score, 
respectively, over all items in \seenitemsk.

Our strategy for constructing the recommendation list \ilistk is to combine relevance and diversity into one score.
For each item \anitem with relevance $\relscore_\anitem$ and diversity $\idiversity_\anitem$,  
we compute the combined score $\combinedscore_\anitem$ by adopting the Clayton copula function~\cite{clayton_copula}
\begin{equation}
    \combinedscore_\anitem = \big[ \normrelscore_i^{-\alpha} +  \normsumdiversity_i^{-\alpha} - 1 \big]^{-1/\alpha},
\label{eq:clayton_copula}
\end{equation}
where $\alpha > 0$ is a regularization parameter.
The list \ilistk is then formed by selecting the top-\ilistsize items from \candidatesk 
according to their combined score $\combinedscore_\anitem$.

We refer to this strategy as \ouralgo.
When the distance diversity function is used we refer to it as \ouralgodistance, and when coverage diversity is used we refer to it as \ouralgocover. A final word on the justification of using the copula function~(\ref{eq:clayton_copula}).
Copulas are functions able to model the cumulative joint distribution of uniform marginal distributions. In general, they are used to represent correlation and dependencies of high-dimensional random variables~\cite{Wang2019NeuralGC, pmlr-v216-peng23a, zeng2022neural, Novianti_2021}.
%However, we here exploit them in a different way. To the best of our knowledge, this is the first attempt to use copulas in this context. 
% Figure~\ref{fig:copula} shows the shape of the Clayton copula in the interval [0, 1], for different values of the $\alpha$ parameter.
The Clayton copula function approaches 1 when both the input variables $u, v$ are maximized, and it is minimized when either of them is 0. The $\alpha$ parameter governs the steepness and folding of the surface: the higher the value of $\alpha$, the more stooped the function is when $u=v$. 

The complexity of the algorithm is discussed in the Appendix.

\section{Experiments}
\label{sec:experiments}

In this section, we assess the performance of our recommendation strategy, either \ouralgodistance or \ouralgocover, in balancing accuracy and diversity. We also compare its effectiveness with several state-of-the-art competitors within the proposed knowledge-exploration framework.

\iffalse
We design our experimental evaluation so as to address the following questions:

\spara{RQ1}: What is the trade-off between accuracy and diversity exhibited by our recommendation strategy, \ouralgo? 

\para{RQ2}: How does our recommendation strategy, \ouralgo, performs compared with state-of-the-art baselines?
    % \item \textbf{RQ3}: At which extent does the $\alpha$ parameter influence the final results?

\para{RQ3}: What is the effect of the relevance component in \ouralgo?
\fi

\subsection{Datasets}
\iffalse
\begin{table*}[t]
    \centering
    \caption{Dataset statistics and Jaccard distances with respect to users and categories.}
    \vspace{-4mm}
    \begin{tabular}{l rrr  rrrr rrrr}
    \toprule
         & & & & \multicolumn{4}{c}{Users-Distance}
         & \multicolumn{4}{c}{Categories-Distance} \\
         \cmidrule(lr){5-8} 
         \cmidrule(lr){9-12}
          Dataset & \multicolumn{1}{c}{\#Users} & \#Items & \#Ratings & \textit{Mean} & \textit{Median} & \textit{Min} & \textit{Max} & \textit{Mean} & \textit{Median} & \textit{Min} & \textit{Max} \\ 
         % \cmidrule(lr){1-1}
         \midrule 
         Movielens-1M & 6\,040 & 3\,706 & 1\,000\,208 & 0.97 & 0.98 & 0.00 & 1.00 & \textbf{0.83} & 1.00 & 0.00 & 1.00 \\
         Coat & 290 & 300 & 6\,960 & 0.97 & 0.99 & 0.00 & 1.00 & \textbf{0.73} & 0.67 & 0.00 & 1.00 \\
         KuaiRec-2.0 & 1\,411 & 3\,327 & 4\,676\,570 & \textbf{0.35} & 0.31 & 0.00 & 0.79 & 0.91 & 1.00 & 0.00 & 1.00\\
         Netflix-Prize & 4\,999 & 1\,112 & 557\,176 & 0.95 & 0.97 & 0.00 & 1.00 & \textbf{0.83} & 0.86 & 0.00 & 1.00\\
         Yahoo-R2 & 21\,181 & 3\,000 & 963\,296 & 0.99 & 1.00 & 0.00 & 1.00 & \textbf{0.26} & 0.00 & 0.00 & 1.00\\
         % \midrule
    \bottomrule
    \end{tabular}
    \label{tab:datasets_statistics}
\end{table*}

\fi
\begin{table}[t]
    \centering
    \caption{Dataset statistics and mean Jaccard distances with respect to users ($\hat{D}_U$) and categories ($\hat{D}_C$).}
    \vspace{-4mm}
    \resizebox{.7\columnwidth}{!}{\begin{tabular}{l rrr r r}
    \toprule
         % & & & & \multicolumn{1}{c}{Mean Users-Distance}
         % & \multicolumn{1}{c}{Mean Categories-Distance} \\
         % \cmidrule(lr){5-8} 
         % \cmidrule(lr){9-12}
          Dataset & \multicolumn{1}{c}{|$\mathcal{U}$|} & {|$\mathcal{I}$|} & \#Ratings & $\hat{D}_U$ & $\hat{D}_C$ \\ 
         % \cmidrule(lr){1-1}
         \midrule 
         Movielens-1M & 6\,040 & 3\,706 & 1\,000\,208 & 0.97 & \textbf{0.83}  \\
         Coat & 290 & 300 & 6\,960 & 0.97 & \textbf{0.73}  \\
         KuaiRec-2.0 & 1\,411 & 3\,327 & 4\,676\,570 & \textbf{0.35} & 0.91 \\
         Netflix-Prize & 4\,999 & 1\,112 & 557\,176 & 0.95 & \textbf{0.83} \\
         Yahoo-R2 & 21\,181 & 3\,000 & 963\,296 & 0.99 & \textbf{0.26} \\
         % \midrule
    \bottomrule
    \end{tabular}}
    \label{tab:datasets_statistics}
\end{table}
We use five benchmark datasets, freely available online.
We ensure that all datasets have category information, 
which is used by our diversity measures.

\spara{Movielens-1M}~\cite{movielens}:\footnote{\url{https://grouplens.org/datasets/movielens/1m/}}
A popular dataset with movie ratings in the range $[1, 5]$, and movie genres.

\spara{Coat}~\cite{coat}:\footnote{\url{https://www.cs.cornell.edu/~schnabts/mnar/}} 
Ratings on coats in the range $[1, 5]$, 
and information on coats' properties. 

\spara{KuaiRec-2.0}~\cite{gao2022kuairec}:\footnote{\url{https://kuairec.com/}}
A recommendation log from a video-sharing mobile app.
Context information is provided, such as \textit{play duration}, \textit{video duration}, and \textit{watch ratio}.
We convert the watch ratios into ratings by interpolating the values from $[0, 2]$ to $[1, 5]$, 
where 0 represents ``never watched'' and 2 represents ``watched twice.'' 
We use the \textit{small} version of the dataset.

\spara{Netflix-Prize}~\cite{Bennett2007TheNP}:\footnote{\url{https://www.kaggle.com/datasets/rishitjavia/netflix-movie-rating-dataset}}
Movie ratings in the range $[1, 5]$. 
We adopt a smaller sample of the original dataset by randomly selecting 
5\,000 items and discard the users with less than 20 interactions. 
Movie categories are acquired from a dataset using the IMDB database.%
\footnote{\url{https://github.com/tommasocarraro/netflix-prize-with-genres}}.

\spara{Yahoo-R2}:\footnote{\url{https://webscope.sandbox.yahoo.com/catalog.php?datatype=i&did=67}}
Song ratings in the range $[1, 5]$. 
Each item is accompanied by artist, album, and genre information. 
We randomly sample 3\,000 items and discard users with less than 20 interactions.

\iffalse
Two considerations should be made regarding the choice of the datasets to be compliant with our methodology: ($i$) they must contain items' information that can be interpreted as categories, and ($ii$) present a suitable items' distance distribution (either in terms of users or categories).
\fi 

\smallskip
Table~\ref{tab:datasets_statistics} provides a summary of the dataset properties, which include the number of users (|$\mathcal{U}$|), the number of items (|$\mathcal{I}$|), the number of ratings (\#Ratings), and the distribution of item distances, calculated based on either users or categories. 
During our experiments, we use Equation~(\ref{equation:distance}) with the distance that exhibits the lowest mean for each dataset. This approach helps us avoid potential bias from large distance values, which could otherwise hinder the effectiveness of the approaches.

% \aris{What does the last sentence mean?
% And why do we have to \textbf{choose} one of the two distances?
% Need to be explained better.}
% \erica{We "choose" the most suitable kind of diversity for the experimental evaluation (i.e., a diversity measure according to which the items are not all distant to each other (e.g., Users-Distance on Movielens-1M). I tried to specify it better.}

\iffalse
Table~\ref{tab:weibull_diversity_upper} further reports the maximum scores for every dataset, in terms of \divdist and \divcov, by varying \expectation{\mathrm{steps}}. To estimate them, we compute the following:
\begin{equation}
    \expectation{\diversity} = \sum_{\iter=1}^{\infty} M_{\iter} \Big( q^{\iter^\gamma} - q^{(\iter+1)^\gamma} \Big)
\end{equation}
where $\diversity \in \{\divdist, \divcov\}$ and $M^D_{\iter}$ is the maximum score obtainable at step \iter.

\aris{I do not understand the last paragraph.
What is \expectation{x}? Do you mean \expectation{steps}? 
And what is this formula? 
I also do not understand what is the purpose of Table 2.
Do we really need it?
Also, what does ``expected maximum'' mean?
}
\erica{Yes, it is \expectation{steps}. \textbf{Important. Giuseppe suggested to use something different than "steps" since in equation 8 we use in the continuous case.} The formula represents an upper bound (in expectation) in terms of scores (either distance or coverage). Table 2 is not necessary here (can go in the appendix), but I think it should go somewhere since in the results table (Table 3), we put also the Delta wrt to the expected maximum scores.}
\fi

\subsection{Competing recommendation strategies}

We evaluate our recommendation algorithm, \ouralgo, 
against the following baseline and state-of-art strategies
that have been designed for the task of increasing diversity in recommender systems.

\spara{Relevance}: This approach recommends the \ilistsize most relevant items, making it a fundamental baseline. Since this strategy is solely focused on maximizing relevance, it represents the most straightforward and basic diversity method, and any other approach must outperform it to be deemed effective.
%It suggests the \ilistsize most relevant items, thus being a reference \textit{baseline} for measuring diversity.  The rationale is that since this strategy is optimized for maximizing relevance only, it is the most "naive" competitor in terms of diversity, and any approach has to perform better in order to be considered. 

\spara{Maximal marginal relevance (MMR)}~\cite{mmr}: 
A classic method used to balance relevance and diversity, 
performed by optimizing the following marginal relevance: 
    \begin{equation*}
        \mbox{MMR} = \argmax_{i \notin L} \left\{ \beta\, \relscore(\auser, \anitem) - (1 - \beta)  \max_{j \in L} \textbf{S}_{i,j} \right\},
    \end{equation*}
    where $\textbf{S}_{i,j} = 1 - \distance(i,j)$. 
    In our experiments, we set $\beta = 0.5$.
    % in order to obtain the perfect trade-off between relevance and diversity.
    
\spara{DUM}~\cite{dum}: This strategy aims at diversifying the suggestions by performing the following diversity-weighted utility maximization:
    \begin{equation*}
        \mbox{DUM} = \argmax_{L \in \Pi} \sum_{h=1}^k \left[f\left(L_{[:h]}\right) - f\left(L_{[:h-1]}\right)\right] \relscore(u,i_h),
    \end{equation*}
    where $\Pi$ denotes all possible permutations of $L$, $L_{[:h]}$ represents the list up to the $h$-th element $i_h$, and 
    $f(X) = \sum_{c \in \categories} \mathbbm{1} \{ \text{exists } i \in X : {i} \text{ covers category } {c}\}$ is the number of categories in $X$. 
    Hence, the function maximizes the relevance of the recommended items weighted by the increase in their coverage.
    
\spara{DPP}~\cite{dpp}: This method utilizes \emph{determinantal point processes} and maximizes diversity 
by iteratively selecting the item $i$ that maximizes the determinant 
of the item-item similarity matrix \textbf{S} defined on a subset of items:
    \begin{equation*}
        \mbox{DPP} = \argmax_{i \notin L} \left\{ \log \det(\textbf{S}_{L \cup \{i\}}) - \log \det(\textbf{S}_{L}) \right\}.
    \end{equation*}

\spara{DGREC}~\cite{dgrec}: 
A \emph{graph neural network} (GNN) based recommender that aims at finding 
a subset of diverse neighbors as well as maximizing the coverage of categories, 
by optimizing the loss function:
    \begin{equation*}
        \mathcal{L}_{\mbox{\footnotesize DGREC}} = 
            \sum_{(u,i) \in E} w_{\textbf{y}_i} \mathcal{L}_{\mbox{\footnotesize BPR}}(u,i,j) + \lambda ||\Theta||_2^2 ,
    \end{equation*}
    where $w_{\textbf{y}_i}$ is the weight for each sample based on its category, 
    $\lambda$~is a regularization factor, and 
    $\mathcal{L}_{\mbox{\footnotesize BPR}}$ is the Bayesian personalized ranking loss~\cite{bpr} .

\smallskip
Notably, these competitors exhibit significant heterogeneity both in terms of the approaches they employ as well as the specific diversity functions they aim to optimize.

\subsection{Experimental setting}
\label{sec:settings}

To evaluate the performance of the examined recommendation strategies, we divide user interactions into a training and a test set, following an 80-20\% split ratio. When evaluating the accuracy, we only focus on the recommendation list generated in the initial exploration step. Regarding diversity, we consider the complete set of recommendation lists produced across all exploration steps. Our approach also assumes that the entire item catalog is accessible to every user during the simulation. 

To calculate the relevance score $\relscore(\auser, \anitem)$, we employ a black-box model in the form of a neural network based on matrix factorization~\cite{matrix_factorization}. We fine-tune the latent factors of this model for each dataset. For \ouralgo, we use a value of $\alpha = 0.5$ in the Clayton copula. Additionally, we conduct hyperparameter tuning for this parameter, and it appears that it has no significant impact on the results (further details can be found in the Appendix).

We keep the length of the recommendation list, \ilist, fixed at 10, and vary the expected number of steps, \expectation{\mathrm{steps}}, in the range of [5, 10, 20]. This allows us to devise a suitable value for the Weibull parameter $\lambda$ to be used in the simulation experiments, according to Equation~(\ref{eq:expected-steps}). To assess recommendation quality, we use standard metrics: \textit{Hit-Ratio} (\textit{HR}), \textit{Precision}, and \textit{Recall}.
Our experimental results are the average of 20 independent trials, and we use the ANOVA test~\cite{anova} to evaluate statistical significance. The code used in these experiments is made publicly accessible.\footnote{\url{https://github.com/EricaCoppolillo/EXPLORE}}

% \aris{
% \textbf{Important comment 2:} 
% We should not use the notation \expectation{\divdist} and \expectation{\divcov}.
% Usually the notation \expectation{\cdot} is used for analytical expressions. 
% When we have \textbf{observed mean values} estimated from experiments it is better to use a single letter notation, for example $\bar{D}$ and $\bar{C}$.
% And we should refer to these as \textbf{mean values} and not \textbf{expected values}.
% }
% \erica{Ok, I'll fix it.}

% \aris{
% \textbf{Important comment 3:}
% We write ``performance'' to refer to ``standard recommendation quality measures,''
% like precision, recall, etc.
% We also use the term ``accuracy'' sometimes for all these measures collectively, 
% which is not correct, and furthermore we do not do it consistently.
% I feel that the whole discussion is rather confusing and we should be more clear. 
% Also the term ``performance'' is not good, because it is not clear to the reader what we mean.
% I typically use performance for scalability.
% I changed some of occurrences of ``performance'' to  ``recommendation quality,''
% but also not consistently. 
% I suggest to use ``recommendation-quality measures,'' unless we refer 
% explicitly to a measure, precision, recall, etc.
% Or perhaps you have better suggestions ...
% }
% \erica{I'll change them.}

% \aris{For next paper, please use macros for datasets, baselines, and measures
% as it would be much easier to change the notation or font style.}
% \erica{For sure, sorry.}

\subsection{Results}
\label{sec:results}
\label{sec:trade-off}
\begin{figure*}[t]
    \centering
    \begin{subfigure}[b]{\linewidth}
        \centering\includegraphics[width=.8\columnwidth]{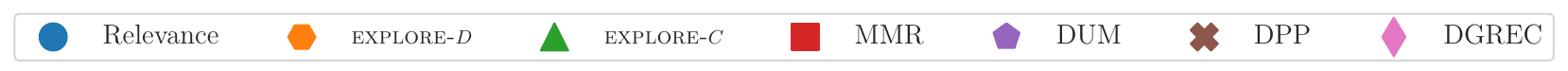}
    \end{subfigure}
    % \medskip
    % \begin{subfigure}[b]{0.33\textwidth}
    %     \includegraphics[width=\columnwidth]{fig/recall/k_10/quality_distance_k_10_users_budget_10_movielens-1m.pdf}
    % \end{subfigure}
    % \begin{subfigure}[b]{0.33\textwidth}
    %     \includegraphics[width=\columnwidth]{fig/recall/k_10/quality_distance_k_10_users_budget_10_coat.pdf}
    % \end{subfigure}
    % \begin{subfigure}[b]{0.33\textwidth}
    %     \includegraphics[width=\columnwidth]{fig/recall/k_10/quality_distance_k_10_users_budget_10_KuaiRec-2.0_small.pdf}
    % \end{subfigure}
    % \medskip
    \begin{subfigure}[b]{0.19\textwidth}
        \includegraphics[width=\columnwidth]{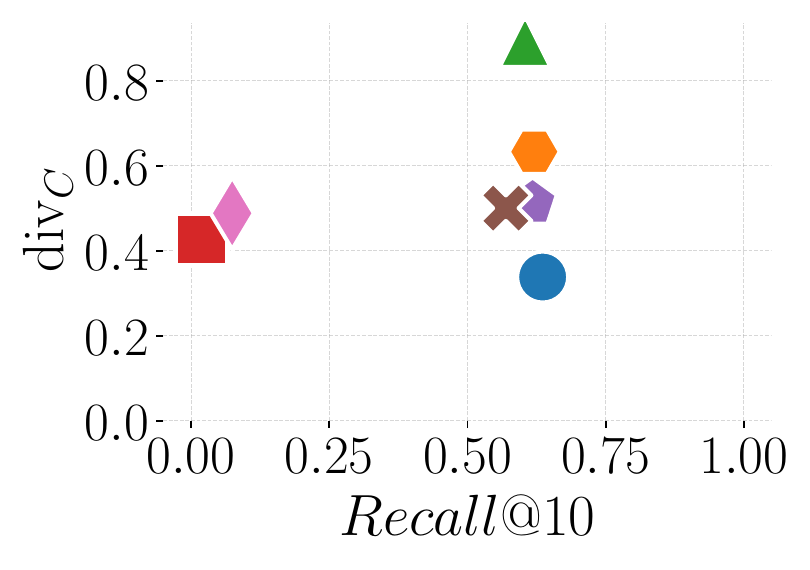}
    \caption{Movielens-1M}
    \end{subfigure}
    \begin{subfigure}[b]{0.19\textwidth}
        \includegraphics[width=\columnwidth]{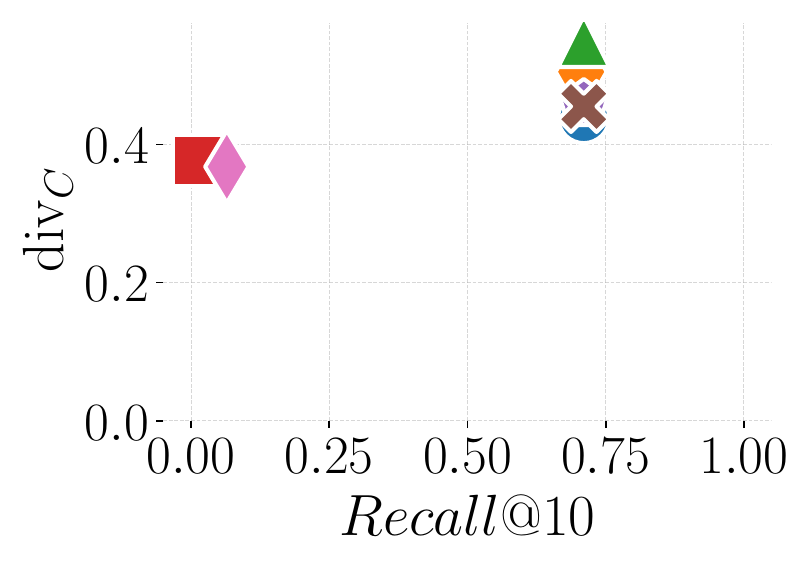}
    \caption{Coat}
    \end{subfigure}
    \begin{subfigure}[b]{0.19\textwidth}
        \includegraphics[width=\columnwidth]{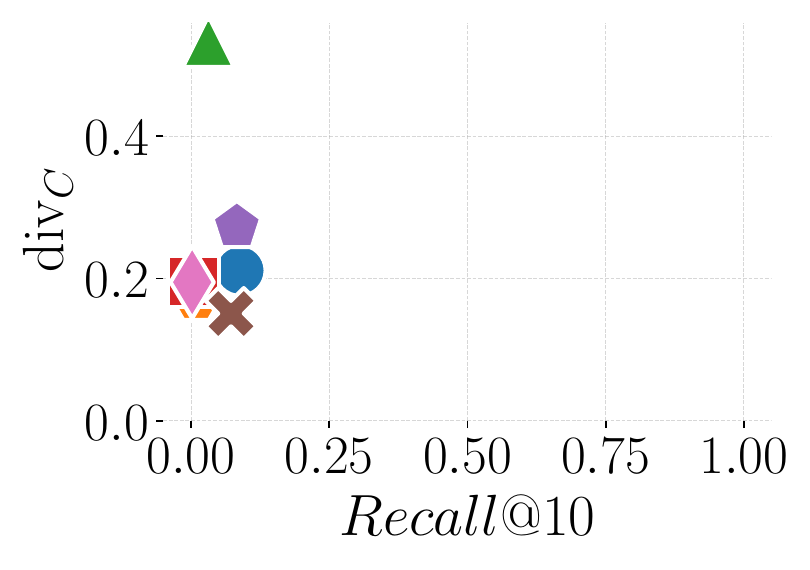}
    \caption{KuaiRec-2.0}
    \label{fig:kuairec_trade-off}
    \end{subfigure}
    % \begin{subfigure}[b]{0.33\textwidth}
    %     \includegraphics[width=\columnwidth]{fig/recall/k_10/quality_distance_k_10_users_budget_10_netflix.pdf}
    % \end{subfigure}
    % \begin{subfigure}[b]{0.33\textwidth}
    %     \includegraphics[width=\columnwidth]{fig/recall/k_10/quality_distance_k_10_users_budget_10_yahoo-r2.pdf}
    % \end{subfigure}
    \begin{subfigure}[b]{0.19\textwidth}
        \includegraphics[width=\columnwidth]{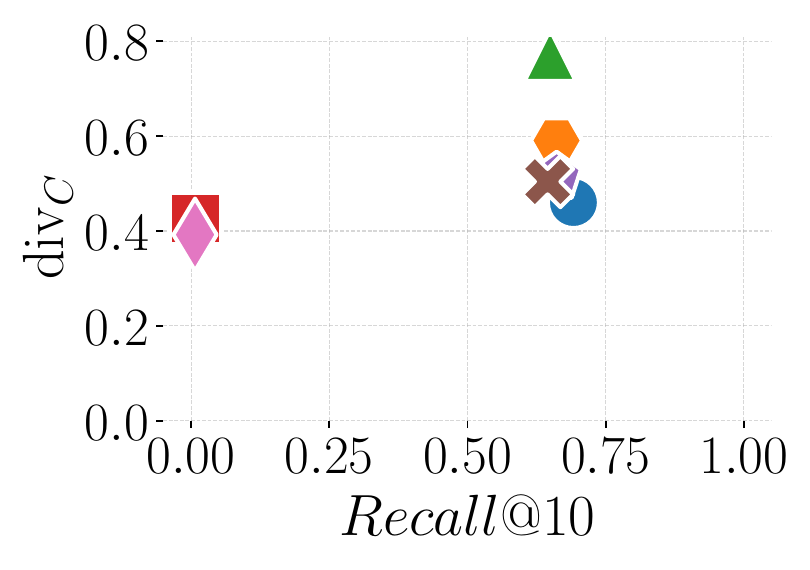}
    \caption{Netflix-Prize}
    \end{subfigure}
    \begin{subfigure}[b]{0.19\textwidth}
        \includegraphics[width=\columnwidth]{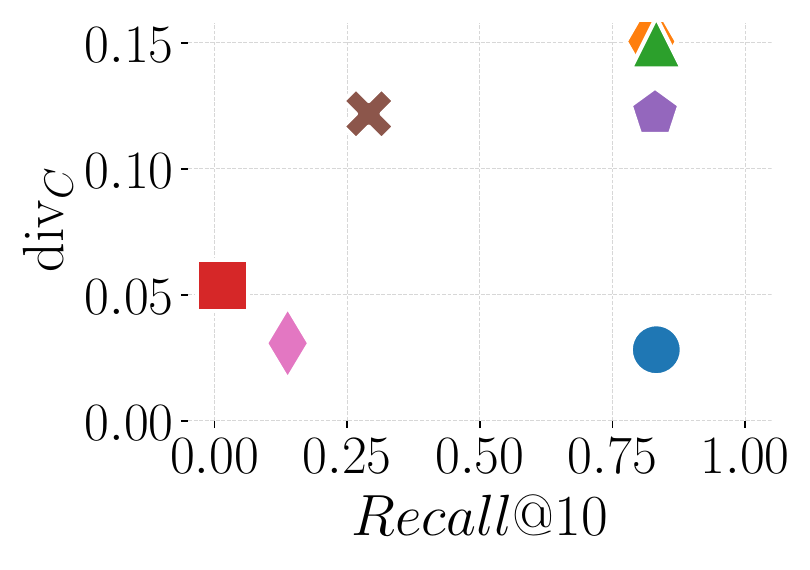}
    \caption{Yahoo-R2}
    \label{fig:kuairec_trade-off}
    \end{subfigure}
    \caption{Trade-off between \divcov and \rm \emph{Recall}@10 \textbf{across all the datasets considered. The X-axis represents recommendation quality, while the Y-axis indicates the diversity score.}}
    \label{fig:trade-off}
\end{figure*}

\spara{Quality-diversity trade-off.}
We initiate our evaluation by assessing the performance of all our strategies in terms of recommendation quality and diversity. Figure~\ref{fig:trade-off} displays the scores for \emph{Recall}@10 (on the $x$-axis) and coverage-based diversity (on the $y$-axis) across all five datasets.
The figure shows that on all datasets, MMR and DGREC exhibit notably poor performance with respect to \textit{Recall}@10. In contrast, the other strategies achieve significantly higher scores, with the Relevance baseline performing the best, which aligns with our expectations.
% Let us focus on the upper row first, in which diversity is computed in terms of distance. Here, the competitors achieve similar results on Movielens-1M and Coat, while behaving differently on KuaiRec-2.0: in the formers, \textit{MMR} and \textit{DGREC} perform slightly worse than the baseline, while \textit{DUM} and \textit{DPP} slightly better; in the latter, \textit{DUM} basically obtains the same score of the baseline, while \textit{DUM}, \textit{MMR} and \textit{DGREC} perform better. 

% Notably, indeed, our strategy \ouralgodistance\ obtains the highest distance score over all the datasets,  decreasing quality to a negligible extent on Movielens-1M and Coat, while degrading more on KuaiRec-2.0. This one, however, appears to be the most problematic in terms of quality, since all the strategies (baseline included) produce significantly lower-quality results.
 
% Similar considerations can be made by analyzing the bottom row, in which the diversity is computed in terms of coverage. Here, the competitors obtain a score similar to the baseline over all the datasets.
In terms of diversity, our method, \ouralgocover, clearly outperforms the other strategies. It achieves a substantially higher diversity score while still delivering relevant recommendations. In fact, it strikes the best trade-off between diversity and relevance. Similar results, both in terms of diversity measures and other evaluation metrics, can be found in the Appendix for reference.

\begin{table}
\centering 
\caption{Diversity scores for $\expectation{\mathrm{steps}} = 5$. Any best scores with a statistical significance $p < 0.05$ are highlighted in bold.}
\vspace{-3mm}
\resizebox{.8\columnwidth}{!}{\begin{tabular}{c c c c c c c c}
\toprule
\multicolumn{1}{c}{Dataset} & Strategy & \multicolumn{1}{c}{$\mathrm{div}_{\!_D}$} & \multicolumn{1}{c}{$\mathrm{div}_{\!_C}$} & \multicolumn{1}{c}{$\steps$} &\multicolumn{1}{c}{$\Delta_{\expecteddivdist}$} & \multicolumn{1}{c}{$\Delta_{\expecteddivcov}$} &\multicolumn{1}{c}{$\Delta_{\mathbb{E}[\mathrm{steps}]}$} \\ 
 \midrule 
\multirow{7}{*}{\rotatebox{90}{Movielens-1M}} & \multirow{1}{*}{Relevance} & 3.67 & 0.22 & 5.0 & 0.27 &0.73 &0.0 \\ 
\cmidrule(lr){2-8}
&\multirow{1}{*}{\ouralgodistance} & \textbf{4.91} & 0.36 & 4.98 & \textbf{0.03} &0.55 &0.0 \\ 
&\multirow{1}{*}{\ouralgocover} & 4.43 & \textbf{0.71} & 4.71 & 0.12 &\textbf{0.11} &0.06 \\ 
\cmidrule(lr){2-8}
&\multirow{1}{*}{MMR} & 3.96 & 0.29 & 4.57 & 0.22 &0.64 &0.09 \\ 
&\multirow{1}{*}{DUM} & 4.4 & 0.33 & 4.98 & 0.13 &0.59 &0.0 \\ 
&\multirow{1}{*}{DPP} & 4.59 & 0.31 & 4.99 & 0.09 &0.61 &0.0 \\ 
&\multirow{1}{*}{DGREC} & 3.36 & 0.37 & 4.49 & 0.33 &0.54 &0.1 \\ 
\cmidrule(lr){1-8}
\multirow{7}{*}{\rotatebox{90}{Coat}} & \multirow{1}{*}{Relevance} & 3.15 & 0.3 & 4.36 & 0.38 &0.3 &0.13 \\ 
\cmidrule(lr){2-8}
&\multirow{1}{*}{\ouralgodistance} & \textbf{3.48} & 0.34 & 4.16 & \textbf{0.31} &0.2 &0.17 \\ 
&\multirow{1}{*}{\ouralgocover} & 3.36 & \textbf{0.35} & 4.13 & 0.33 &\textbf{0.18} &0.17 \\ 
\cmidrule(lr){2-8}
&\multirow{1}{*}{MMR} & 2.43 & 0.26 & 3.54 & 0.52 &0.39 &0.29 \\ 
&\multirow{1}{*}{DUM} & 3.11 & 0.3 & 4.31 & 0.38 &0.3 &0.14 \\ 
&\multirow{1}{*}{DPP} & 3.28 & 0.31 & 4.33 & 0.35 &0.27 &0.13 \\ 
&\multirow{1}{*}{DGREC} & 2.2 & 0.24 & 3.2 & 0.56 &0.44 &0.36 \\ 
\cmidrule(lr){1-8}
\multirow{7}{*}{\rotatebox{90}{KuaiRec-2.0}} & \multirow{1}{*}{Relevance} & 0.76 & 0.13 & 4.81 & 0.81 &0.74 &0.04 \\ 
\cmidrule(lr){2-8}
&\multirow{1}{*}{\ouralgodistance} & \textbf{1.56} & 0.11 & 3.54 & \textbf{0.61} &0.78 &0.29 \\ 
&\multirow{1}{*}{\ouralgocover} & 1.08 & \textbf{0.34} & 4.09 & 0.73 &\textbf{0.32} &0.18 \\ 
\cmidrule(lr){2-8}
&\multirow{1}{*}{MMR} & 1.25 & 0.12 & 3.89 & 0.68 &0.76 &0.22 \\ 
&\multirow{1}{*}{DUM} & 0.83 & 0.17 & 4.8 & 0.79 &0.66 &0.04 \\ 
&\multirow{1}{*}{DPP} & 1.38 & 0.09 & 4.75 & 0.65 &0.82 &0.05 \\ 
&\multirow{1}{*}{DGREC} & 0.77 & 0.11 & 2.64 & 0.81 &0.78 &0.47 \\ 
\cmidrule(lr){1-8}
\multirow{7}{*}{\rotatebox{90}{Netflix}} & \multirow{1}{*}{Relevance} & 4.04 & 0.32 & 4.86 & 0.2 &0.59 &0.03 \\ 
\cmidrule(lr){2-8}
&\multirow{1}{*}{\ouralgodistance} & \textbf{4.62} & 0.38 & 4.75 & \textbf{0.09} &0.51 &0.05 \\ 
&\multirow{1}{*}{\ouralgocover} & 3.97 & \textbf{0.6} & 4.43 & 0.21 &\textbf{0.22} &0.11 \\ 
\cmidrule(lr){2-8}
&\multirow{1}{*}{MMR} & 3.56 & 0.3 & 4.19 & 0.3 &0.61 &0.16 \\ 
&\multirow{1}{*}{DUM} & 4.16 & 0.36 & 4.89 & 0.18 &0.53 &0.02 \\ 
&\multirow{1}{*}{DPP} & 4.38 & 0.34 & 4.88 & 0.13 &0.56 &0.02 \\ 
&\multirow{1}{*}{DGREC} & 3.0 & 0.26 & 3.74 & 0.41 &0.66 &0.25 \\ 
\cmidrule(lr){1-8}
\multirow{7}{*}{\rotatebox{90}{Yahoo-R2}} & \multirow{1}{*}{Relevance} & 0.66 & 0.02 & \textbf{4.77} & 0.87 &0.77 &\textbf{0.05} \\ 
\cmidrule(lr){2-8}
&\multirow{1}{*}{\ouralgodistance} & \textbf{4.4} & \textbf{0.08} & 4.49 & \textbf{0.13} &\textbf{0.1} &0.1 \\ 
&\multirow{1}{*}{\ouralgocover} & 4.38 & \textbf{0.08} & 4.47 & \textbf{0.13} &\textbf{0.1} &0.11 \\ 
\cmidrule(lr){2-8}
&\multirow{1}{*}{MMR} & 2.45 & 0.04 & 3.96 & 0.52 &0.55 &0.21 \\ 
&\multirow{1}{*}{DUM} & 4.38 & 0.07 & 4.72 & \textbf{0.13} &0.21 &0.06 \\ 
&\multirow{1}{*}{DPP} & 4.38 & 0.07 & 4.72 & \textbf{0.13} &0.21 &0.06 \\ 
&\multirow{1}{*}{DGREC} & 1.02 & 0.02 & 3.68 & 0.8 &0.77 &0.26 \\ 
\bottomrule
\end{tabular}}
\label{tab:partial_scores_table}
\end{table}

\spara{Best performing diversity strategy.}
%Now that we have assessed the recommendation quality of the different strategies, we focus on diversity.
Table~\ref{tab:partial_scores_table} presents a comprehensive analysis of \divdist, \divcov and $\steps$ when $\expectation{\mathrm{steps}} = 5$. Additionally, we report the deviations from the maximum diversity scores in terms of distance and coverage (reported in the Appendix), denoted as $\Delta_{\expecteddivdist}$ and $\Delta_{\expecteddivcov}$, along with $\Delta_{\mathbb{E[\mathrm{steps}]}}$. %Additional details can be found in the Appendix.

%Any highest scores with a statistical significance of $p < 0.05$ have been highlighted in bold. 

% \aris{I suggest to put Table 2 in the appendix, 
% and here say that we have also estimated the maximum possible diversity measure, 
% without considering relevance, and we report the relative distance, 
% and then say that details are in the appendix.
% Also, I would not call the numbers in table 2 as ``expected'' but as ``maximum''.
% }

% The results have been obtained by averaging 20 independent trials and we adopted ANOVA~\cite{anova} to perform the statistical test. 
%%% ARIS: the previous sentence was said before in Settings.

We observe that our strategy, either \ouralgodistance or \ouralgocover, consistently outperforms the competitors in terms of both \divdist and \divcov across all datasets. We also show how these values deviate from the expected maximum values. Notably, on the Movielens-1M dataset, their scores are very close to their maxima. Our strategy achieves significantly higher scores than the competitors on all datasets, especially in terms of coverage.

Regarding the number of steps, as mentioned in Section~\ref{sec:problem}, the relevance plays a fundamental role in our exploration process. Therefore, it is expected that our strategy performs slightly worse than other competitors, particularly the Relevance baseline. Nevertheless, our primary objective is to maximize recommendation diversity while maintaining relevance as high as possible.
%guarda la mail
Additional details on the experiments are in the Appendix.

\begin{table}
 \centering 
\caption{Results obtained in terms of \divdist, \divcov and \rm steps \textbf{by including (}\rm{w}\textbf{) and excluding (}\rm{w/o}\textbf{) relevance from our recommendation strategies. Positive relative changes ($\Delta_\mathrm{w}$) are reported in bold.}}
\vspace{-3mm}
\resizebox{0.62\columnwidth}{!}{\begin{tabular}{c c c c c c}
\toprule
{} & Strategy & Relevance & \multicolumn{1}{c}{$\mathrm{div}_{\!_D}$} & \multicolumn{1}{c}{$\mathrm{div}_{\!_C}$} & \multicolumn{1}{c}{$\steps$} \\ 
 \cmidrule{1-6} 
\multirow{6}{*}{\rotatebox{90}{Movielens-1M}} & \multirow{3}{*}{\ouralgodistance} & w& 9.77 & 0.63 & 9.85 \\ 
& & w/o& 6.66 & 0.53 & 6.73 \\ 
\cmidrule(lr){4-6}
& & $\Delta_\mathrm{w}$ & \textbf{+0.32} & \textbf{+0.16} & \textbf{+0.32} \\ 
\cmidrule(lr){3-6}
& \multirow{3}{*}{\ouralgocover} & w& 8.84 & 0.89 & 9.7 \\ 
& & w/o& 6.56 & 0.86 & 7.0 \\ 
\cmidrule(lr){4-6}
& & $\Delta_\mathrm{w}$ & \textbf{+0.26} & \textbf{+0.03} & \textbf{+0.28} \\ 
\cmidrule(lr){1-6}
\multirow{6}{*}{\rotatebox{90}{Coat}} & \multirow{3}{*}{\ouralgodistance} & w& 6.73 & 0.51 & 8.02 \\ 
& & w/o& 5.5 & 0.47 & 6.41 \\ 
\cmidrule(lr){4-6}
& & $\Delta_\mathrm{w}$ & \textbf{+0.18} & \textbf{+0.08} & \textbf{+0.2} \\ 
\cmidrule(lr){3-6}
& \multirow{3}{*}{\ouralgocover} & w& 6.31 & 0.55 & 7.81 \\ 
& & w/o& 5.18 & 0.49 & 6.34 \\ 
\cmidrule(lr){4-6}
& & $\Delta_\mathrm{w}$ & \textbf{+0.18} & \textbf{+0.11} & \textbf{+0.19} \\ 
\cmidrule(lr){1-6}
\multirow{6}{*}{\rotatebox{90}{KuaiRec-2.0}} & \multirow{3}{*}{\ouralgodistance} & w& 3.06 & 0.17 & 6.86 \\ 
& & w/o& 2.38 & 0.13 & 4.47 \\ 
\cmidrule(lr){4-6}
& & $\Delta_\mathrm{w}$ & \textbf{+0.22} & \textbf{+0.24} & \textbf{+0.35} \\ 
\cmidrule(lr){3-6}
& \multirow{3}{*}{\ouralgocover} & w& 2.22 & 0.53 & 7.95 \\ 
& & w/o& 2.01 & 0.49 & 6.04 \\ 
\cmidrule(lr){4-6}
& & $\Delta_\mathrm{w}$ & \textbf{+0.09} & \textbf{+0.08} & \textbf{+0.24} \\ 
\cmidrule(lr){1-6}
\multirow{6}{*}{\rotatebox{90}{Netflix}} & \multirow{3}{*}{\ouralgodistance} & w& 9.03 & 0.59 & 9.24 \\ 
& & w/o& 7.42 & 0.54 & 7.52 \\ 
\cmidrule(lr){4-6}
& & $\Delta_\mathrm{w}$ & \textbf{+0.18} & \textbf{+0.08} & \textbf{+0.19} \\ 
\cmidrule(lr){3-6}
& \multirow{3}{*}{\ouralgocover} & w& 7.92 & 0.77 & 8.69 \\ 
& & w/o& 6.71 & 0.75 & 7.38 \\ 
\cmidrule(lr){4-6}
& & $\Delta_\mathrm{w}$ & \textbf{+0.15} & \textbf{+0.03} & \textbf{+0.15} \\ 
\cmidrule(lr){1-6}
\multirow{6}{*}{\rotatebox{90}{Yahoo-R2}} & \multirow{3}{*}{\ouralgodistance} & w& 8.71 & 0.15 & 8.73 \\ 
& & w/o& 6.23 & 0.11 & 6.3 \\ 
\cmidrule(lr){4-6}
& & $\Delta_\mathrm{w}$ & \textbf{+0.28} & \textbf{+0.27} & \textbf{+0.28} \\ 
\cmidrule(lr){3-6}
& \multirow{3}{*}{\ouralgocover} & w& 8.67 & 0.15 & 8.7 \\ 
& & w/o& 6.25 & 0.11 & 6.31 \\ 
\cmidrule(lr){4-6}
& & $\Delta_\mathrm{w}$ & \textbf{+0.28} & \textbf{+0.27} & \textbf{+0.27} \\ 
\bottomrule
\end{tabular}}
\label{tab:ablation_table}
\end{table}

\spara{Ablation study.}
In our final investigation, we explore the advantages of combining both relevance and diversity through the copula function in Equation~(\ref{eq:clayton_copula}), in contrast to a simpler strategy that neglects relevance and relies on Equation~(\ref{equation:marginal-diversity}).

Table~\ref{tab:ablation_table} presents a summary of the results obtained for $\expectation{\mathrm{steps}} = 10$. For each strategy, we provide the values for \divdist, \divcov, and actual steps \steps. The scores are computed for two variants: one where relevance is included through the copula function (w) and another where it is ignored (w/o). The table also reports the differences in scores ($\Delta_{\mathrm{w}}$). %Positive variations in adopting quality with respect to considering diversity only are highlighted in bold. 
We can observe that the combination has a positive effect
both in terms of diversity and number of steps.

\section{Conclusion and future work}
\label{sec:conclusion}
In this study, we addressed recommendation diversity by introducing a user-behavior model where relevance drives engagement. We devised a recommendation strategy that optimizes the delivery of diverse knowledge to users in accordance with the underlying user behavior. The experimental analysis confirms that our approach is effective, yet it remains open to further enhancements. First, the behavioral model can be refined to encompass more sophisticated scenarios. These include alternative actions such as refreshing the list or guiding its composition, and incorporating dynamic adjustments to the weariness probability beyond temporal decay. Also, our model assumes that the underlying relevance score captures a user's actual interest in an item. However, the relevance score is computed through an algorithm that is not necessarily totally accurate. We can still adapt the user behavior model to account for such inaccuracy by incorporating a random discount factor associated with the relevance of each item in the list. Finally, the proposed strategy can be improved in several ways: for example, by integrating different distance measures or tailoring it to encompass additional metrics beyond diversity, such as serendipity or fairness.

\begin{comment}
In particular, we ($i$) defined $D(C)$ and $cov(D)$ as metrics to quantify the diversity induced on the final interactions set of users, either at item- or group-level; ($ii$) developed a novel simulation methodology to analyze the user-recommender interaction in the long-term, inducing a proven interplay among \textit{relevance}, \textit{diversity} and \textit{user behavior}; and \textit{(iii)} we proposed two recommendation strategies that account for both items utility and diversity. In the experimental section, we show how our proposal overcomes several state-of-the-art competitors over three different popular datasets. 
In future work, we aim to improve user behavior by incorporating different choices of item picking and assuming dynamic patience probabilities.

\begin{itemize}
    \item "Refresh" button for getting new recommendations in the simulation process
    \item Considering entropy in the strategy
\end{itemize}
\end{comment}

\clearpage
% references are excluded from the 8 main pages 
\bibliographystyle{plainnat}

\newpage
\appendix
\section{Appendix}

\begin{figure*}[t]
    \centering
    \begin{subfigure}[b]{.33\textwidth}
        \includegraphics[width=\columnwidth]{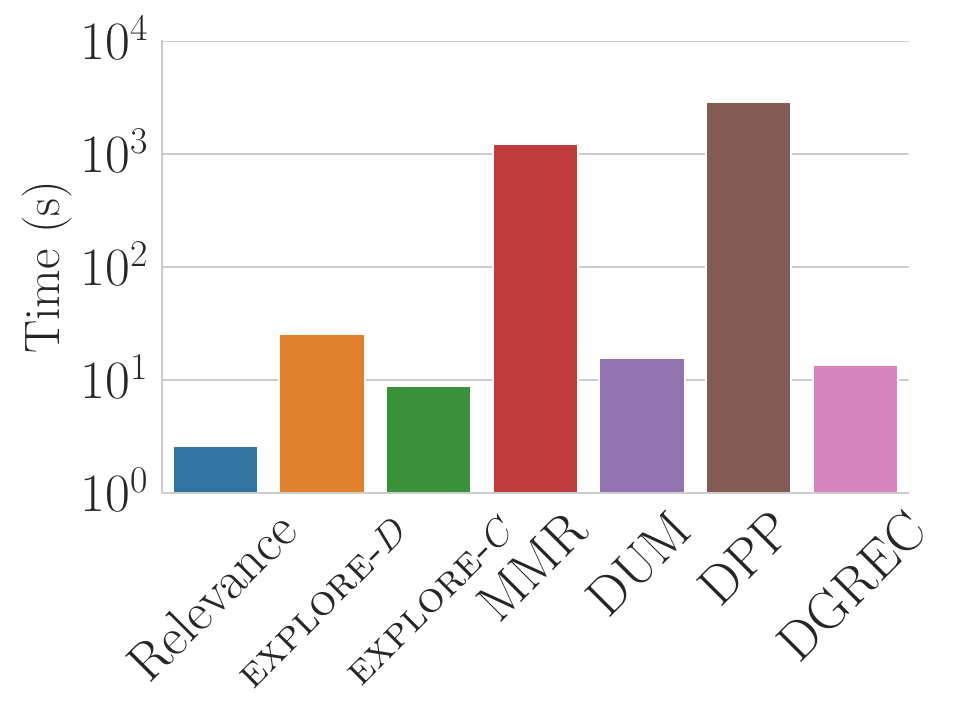}
        \caption{Movielens-1M}
    \end{subfigure}
    \begin{subfigure}[b]{.33\textwidth}
        \includegraphics[width=\columnwidth]{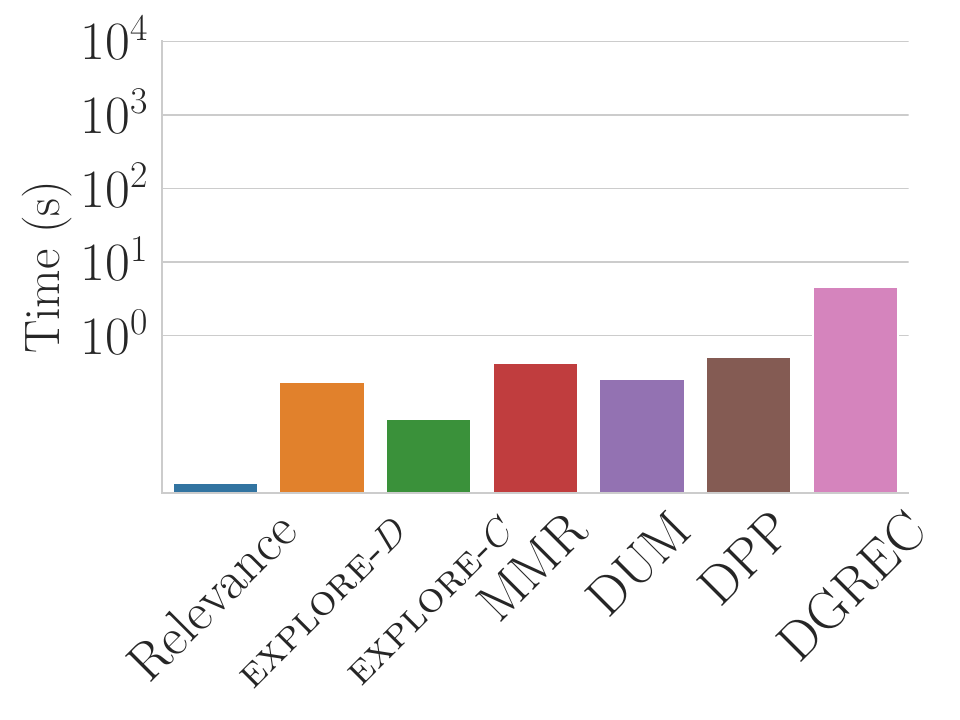}
        \caption{Coat}
    \end{subfigure}
    \medskip
    \begin{subfigure}[b]{.33\textwidth}
        \includegraphics[width=\columnwidth]{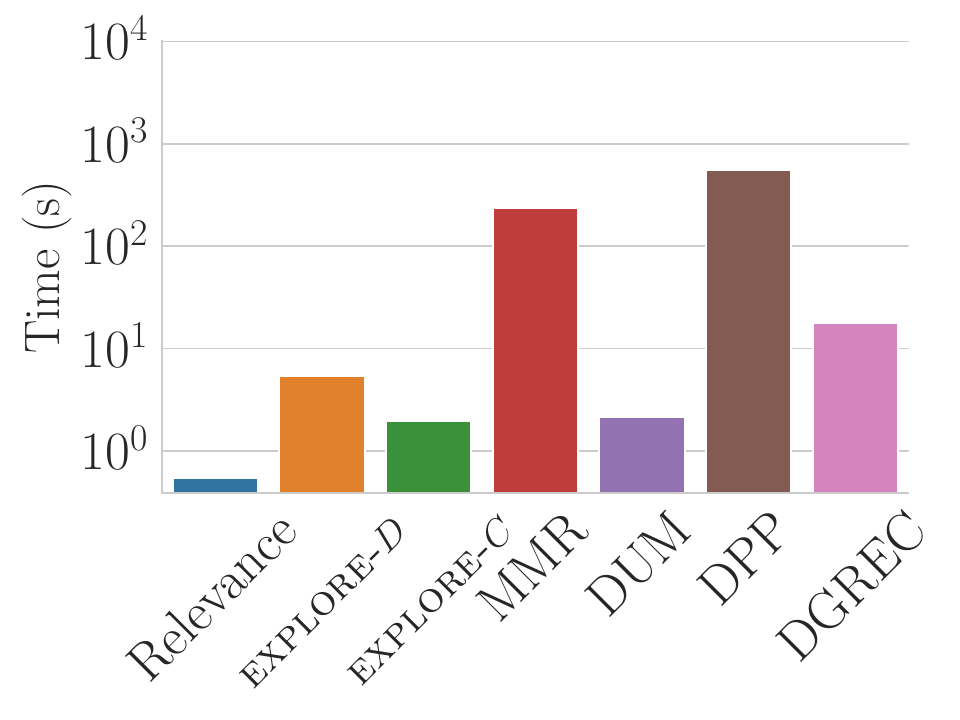}
        \caption{KuaiRec-2.0}
    \end{subfigure}
    \begin{subfigure}[b]{.33\textwidth}
        \includegraphics[width=\columnwidth]{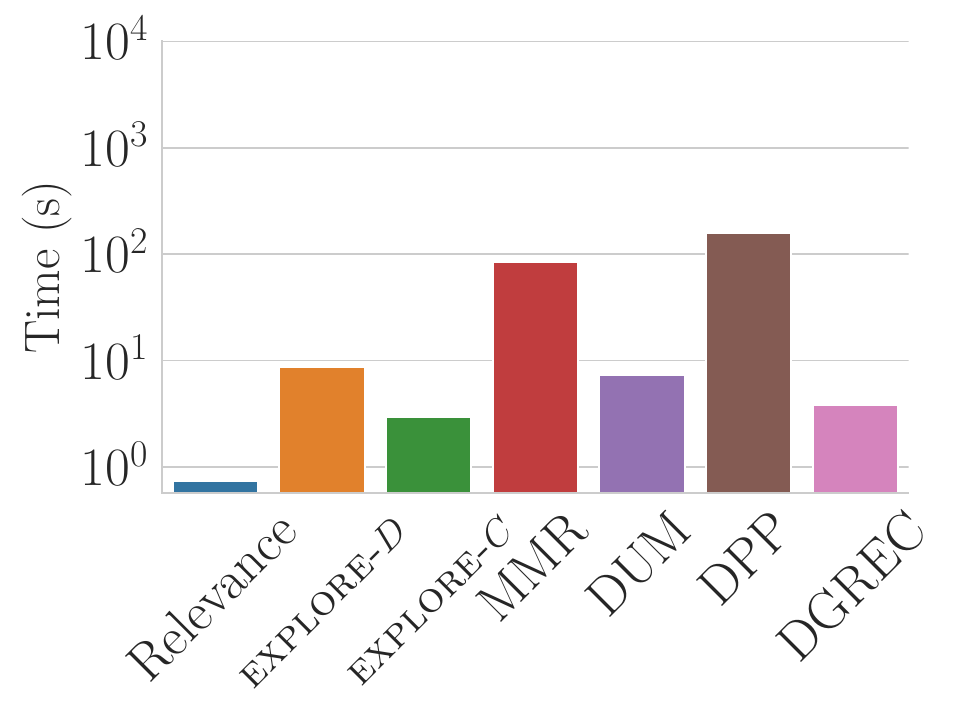}
        \caption{Netflix-Prize}
    \end{subfigure}
    \medskip
    \begin{subfigure}[b]{.33\textwidth}
        \includegraphics[width=\columnwidth]{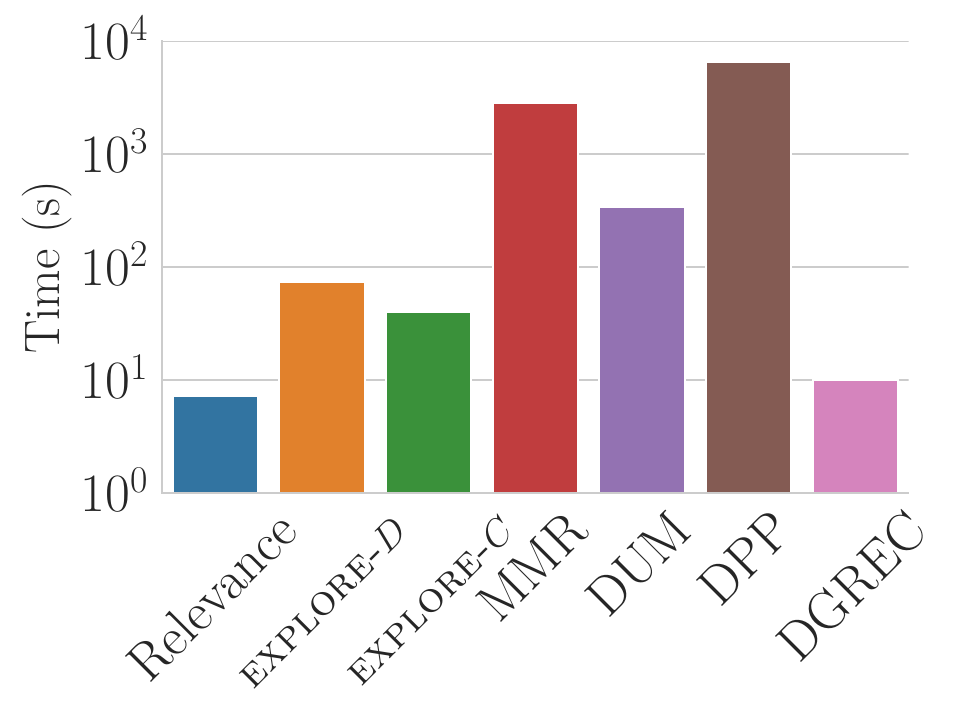}
        \caption{Yahoo-R2}
    \end{subfigure}
    \vspace{-5mm}
    \caption{Timing for producing a recommendation list. The X-axis reports the strategies, while the Y-axis shows the recommendation time (in seconds).}
    \label{fig:times}
\end{figure*}

\spara{Complexity.} In the following, we analyze the computational complexity of the proposed algorithm \ouralgo. Besides the (black-box) recommender system, the critical point is the generation of list \ilistk to be presented to users, where $t$ is an exploration step in a session. Within an online setting, this list has to be dynamically generated and hence can represent a bottleneck. For the generation of \ilistk, we need to consider Equations~\ref{equation:marginal-diversity} and~\ref{eq:clayton_copula}. By considering that \seenitemsk is computed incrementally, the cost of computing $\idiversity_\anitem$ in~\ref{equation:marginal-diversity} is $O(td)$ when adopting \divdist, and $O(|\categories|)$ when adopting \divcov. Here, \textit{d} is the computational cost associated with the Jaccard distance. In total, the worst-case cost for generating a list of \textit{k} elements by considering \noitems items is either $O(\noitems dk^2)$ or $O(\noitems k|\categories|)$.
\\
We can observe the following. (1) The number \noitems of items to consider could be large (in principle, the entire item catalog). However, since $\idiversity_\anitem$ is combined with $\relscore_\anitem$ in Equation~\ref{eq:clayton_copula}, we can filter out low-relevance items, as they will severely affect the value of $\combinedscore_\anitem$ due to the properties of the copula function. Notice also that, in a practical implementation, sampling strategies on portions of the catalog can also be devised to narrow the focus only on a subset of items.
(2) The cost $d$ for computing $d(i, j)$, for two generic items, can be $O(\nousers)$, where \nousers is the total number of users. To relieve this cost, the scores for popular items can be precomputed. Notice that in typical settings the distribution of items is heavy-tailed, thus we can expect that the number of distance scores to precompute is not intractably large.
\\
Figure~\ref{fig:times} shows that indeed our strategy exhibits comparable running times with those of other strategies, while significantly overcoming competitors such as MMR and DPP (that struggle especially with large datasets).

\spara{Experiments.} We present additional experiments regarding the recommendation quality of the strategies, 
the diversity scores, and the run time required to provide the recommendation list.

\begin{figure*}[t]
    \centering
    \begin{subfigure}[b]{\linewidth}
        \includegraphics[width=\columnwidth]{fig/legend.pdf}
    \end{subfigure}
    % \vspace{-5mm}
    \medskip
    \begin{subfigure}[b]{0.3\textwidth}
        \includegraphics[width=\columnwidth]{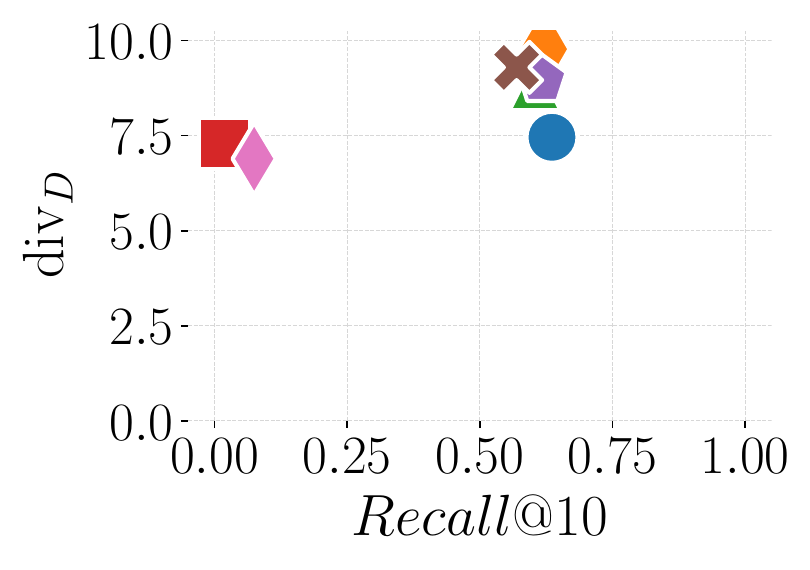}
    \end{subfigure}
    \begin{subfigure}[b]{0.3\textwidth}
        \includegraphics[width=\columnwidth]{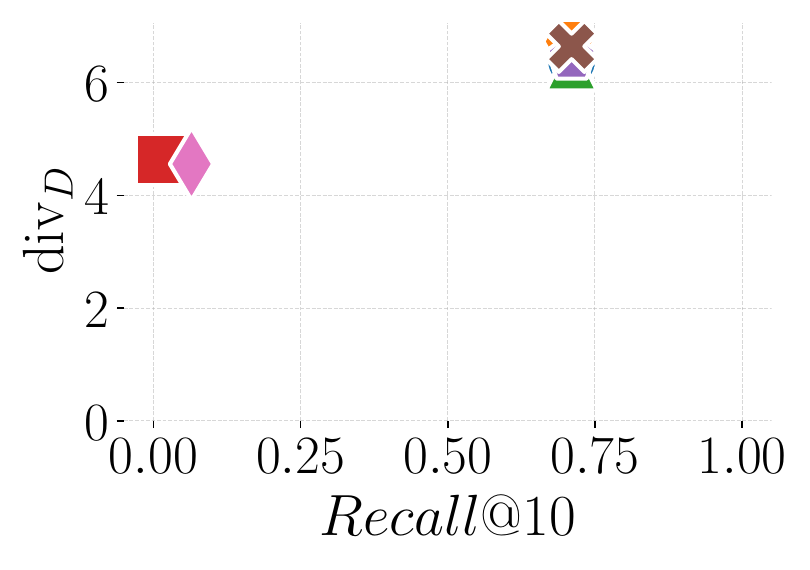}
    \end{subfigure}
    \begin{subfigure}[b]{0.3\textwidth}
        \includegraphics[width=\columnwidth]{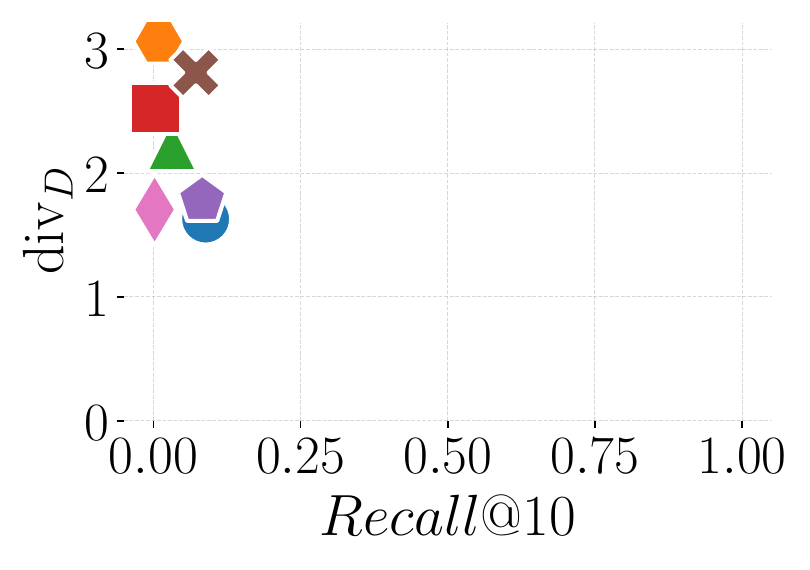}
    \end{subfigure}
    \medskip
    \begin{subfigure}[b]{0.3\textwidth}
        \includegraphics[width=\columnwidth]{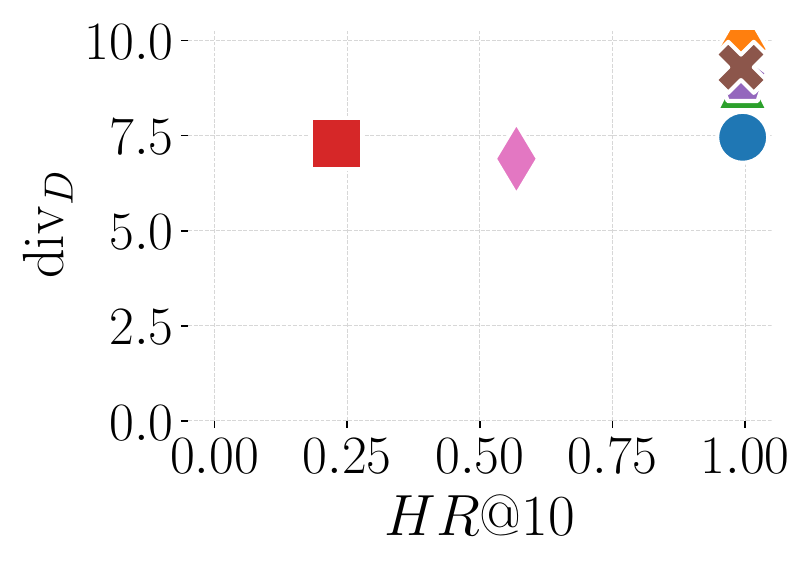}
    \end{subfigure}
    \begin{subfigure}[b]{0.3\textwidth}
        \includegraphics[width=\columnwidth]{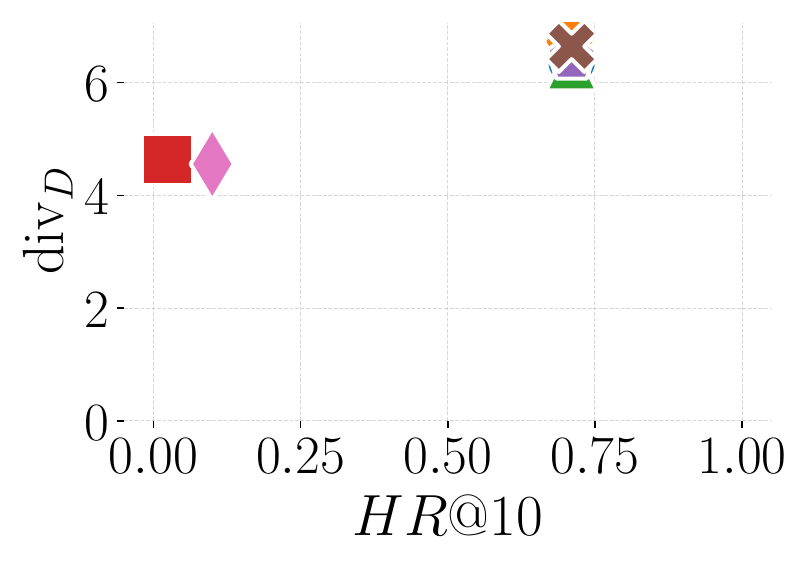}
    \end{subfigure}
    \begin{subfigure}[b]{0.3\textwidth}
        \includegraphics[width=\columnwidth]{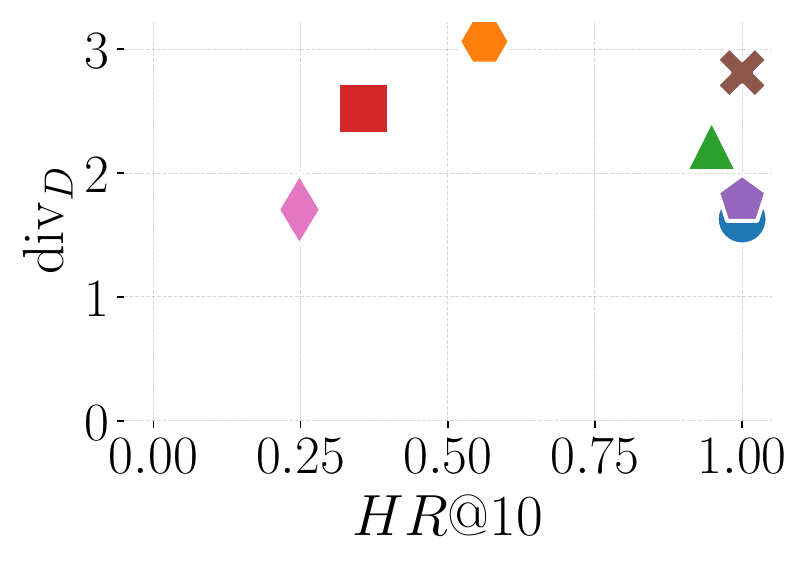}
    \end{subfigure}
    \medskip
    \begin{subfigure}[b]{0.3\textwidth}
        \includegraphics[width=\columnwidth]{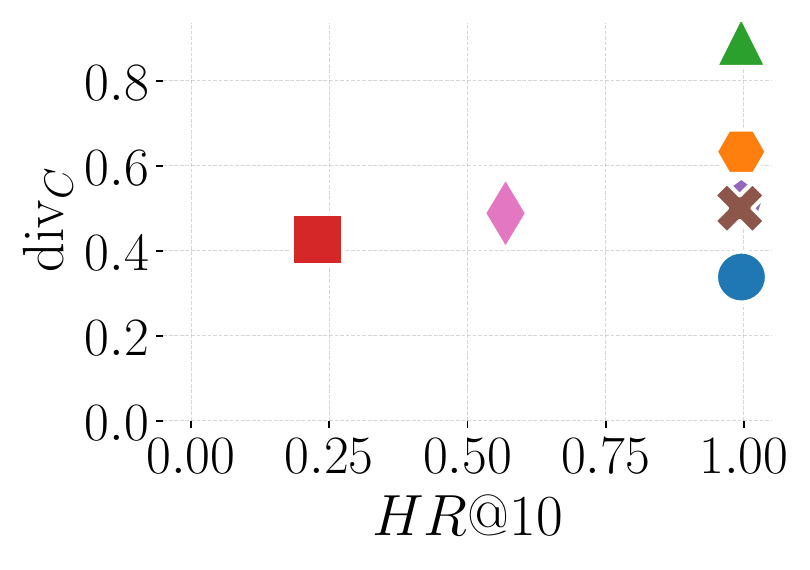}
    % \caption{Movielens-1M}
    \end{subfigure}
    \begin{subfigure}[b]{0.3\textwidth}
        \includegraphics[width=\columnwidth]{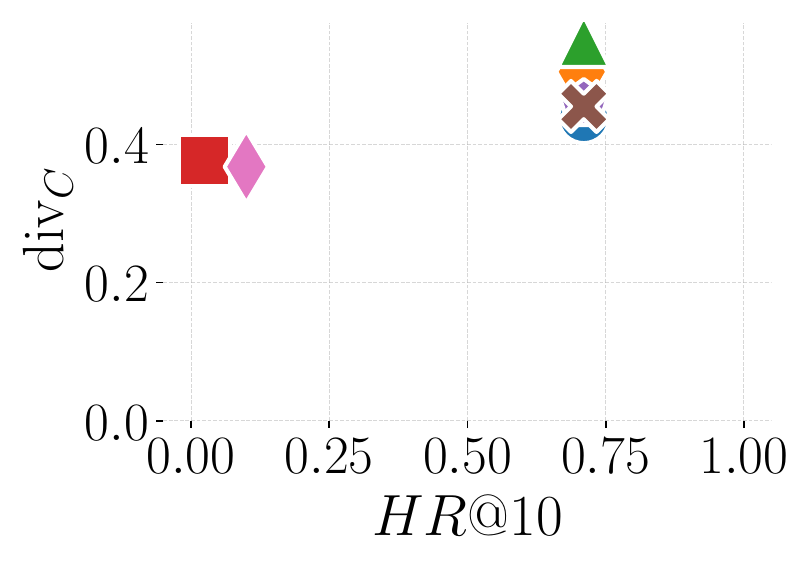}
    \end{subfigure}
    \begin{subfigure}[b]{0.3\textwidth}
        \includegraphics[width=\columnwidth]{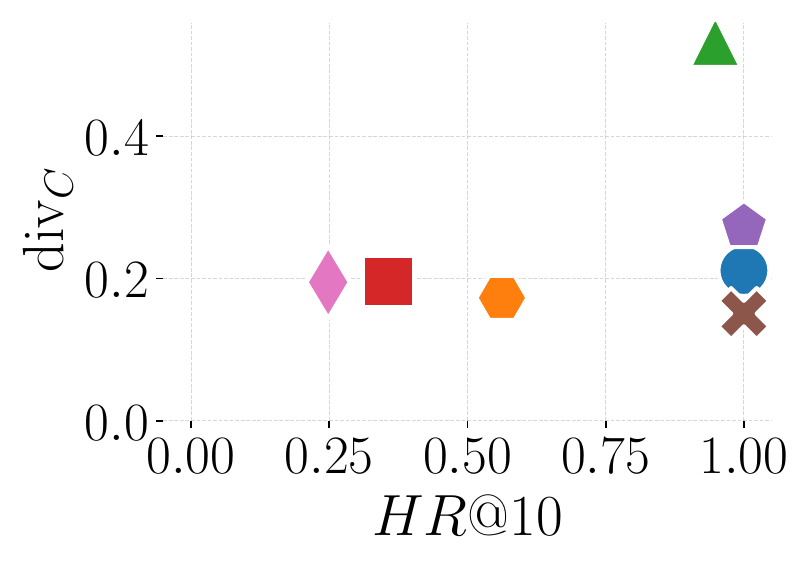}
    % \caption{KuaiRec-2.0}
    \end{subfigure}
    \medskip
    \begin{subfigure}[b]{0.3\textwidth}
        \includegraphics[width=\columnwidth]{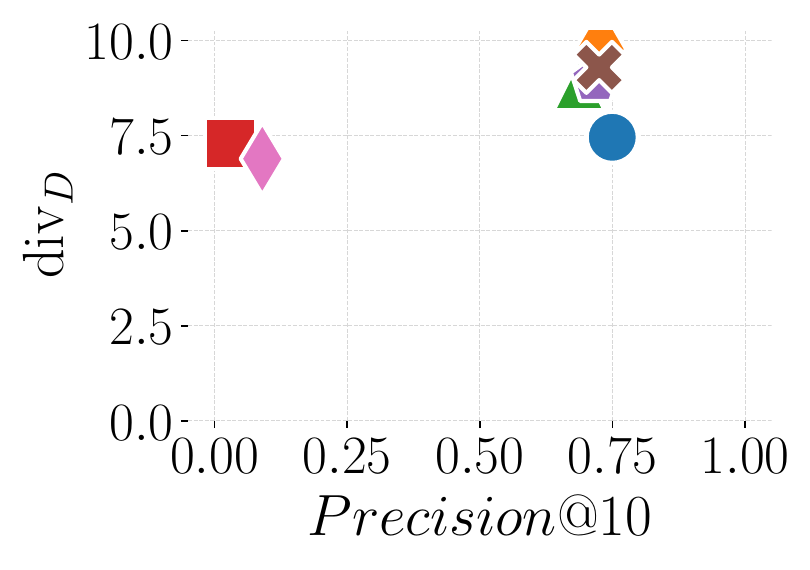}
    \end{subfigure}
    \begin{subfigure}[b]{0.3\textwidth}
        \includegraphics[width=\columnwidth]{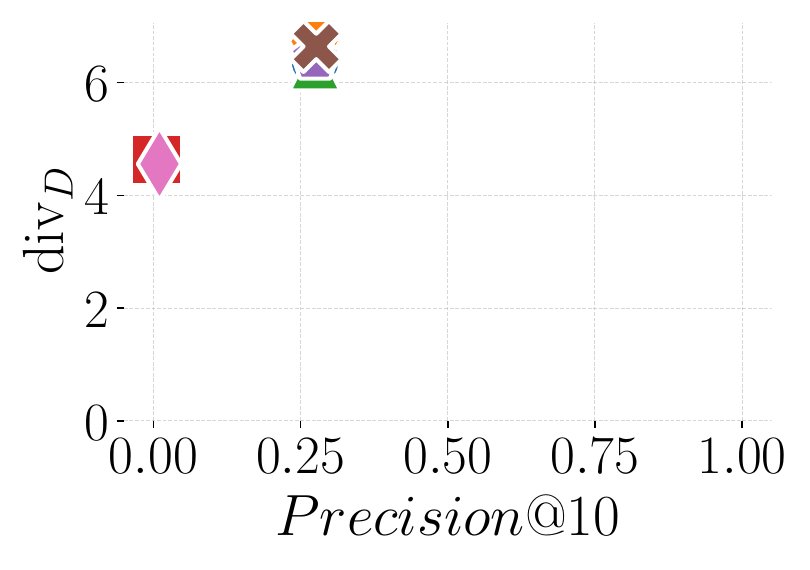}
    \end{subfigure}
    \begin{subfigure}[b]{0.3\textwidth}
        \includegraphics[width=\columnwidth]{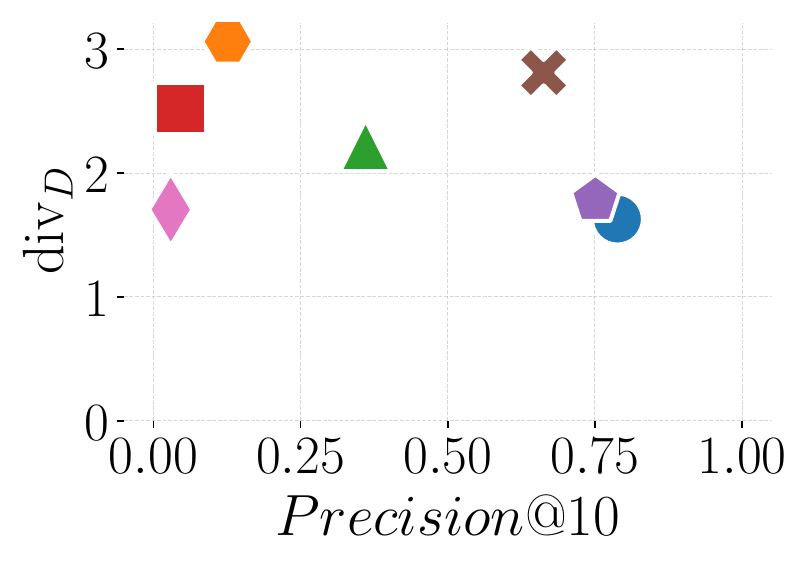}
    \end{subfigure}
    \medskip
    \begin{subfigure}[b]{0.3\textwidth}
        \includegraphics[width=\columnwidth]{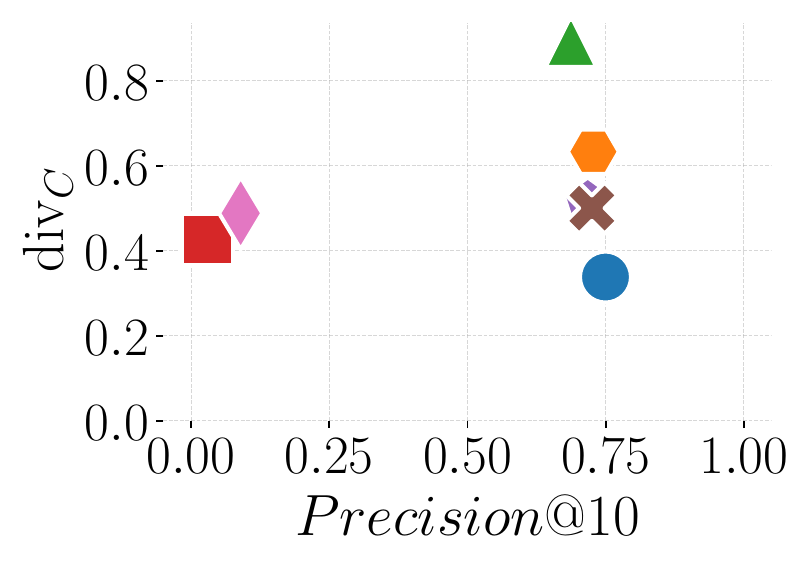}
    \caption{Movielens-1M}
    \end{subfigure}
    \begin{subfigure}[b]{0.3\textwidth}
        \includegraphics[width=\columnwidth]{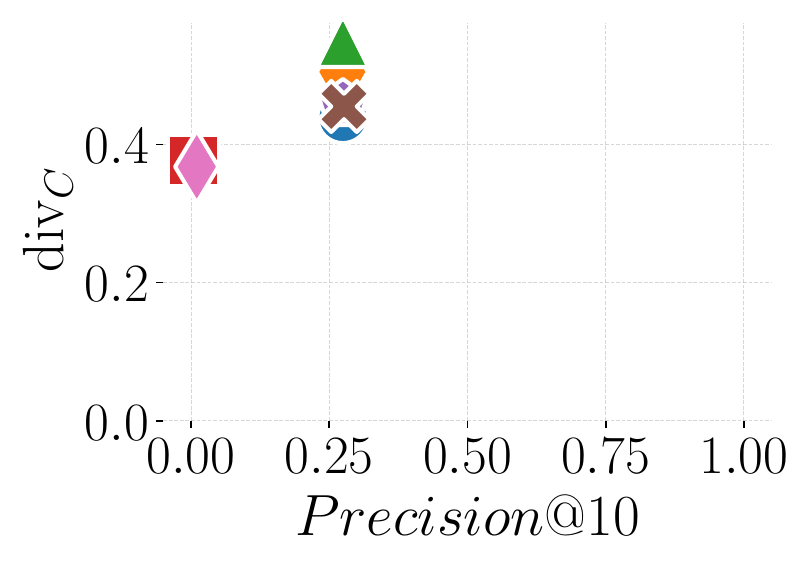}
    \caption{Coat}
    \end{subfigure}
    \begin{subfigure}[b]{0.3\textwidth}
        \includegraphics[width=\columnwidth]{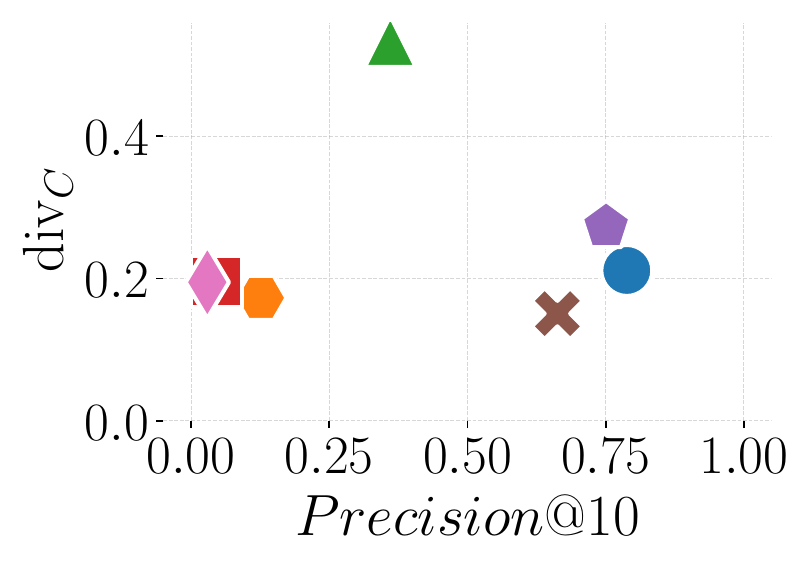}
    \caption{KuaiRec-2.0}
    \end{subfigure}
    \vspace{-5mm}
    \caption{Trade-off between either \divdist or \divcov and either \rm  \emph{Recall}@10, \emph{HR}@10, \textbf{or} \rm \emph{Precision}@10 \textbf{over Movielens-1M, Coat and KuaiRec-2.0. The X-axis shows the recommendation quality while the Y-axis represents the diversity score.}}
    \label{fig:quality-diversity-tradeoff-1}
\end{figure*}

\begin{figure*}[t]
    \centering
    \begin{subfigure}[b]{\linewidth}
        \includegraphics[width=\columnwidth]{fig/legend.pdf}
    \end{subfigure}
     \begin{subfigure}[b]{0.3\textwidth}
        \includegraphics[width=\columnwidth]{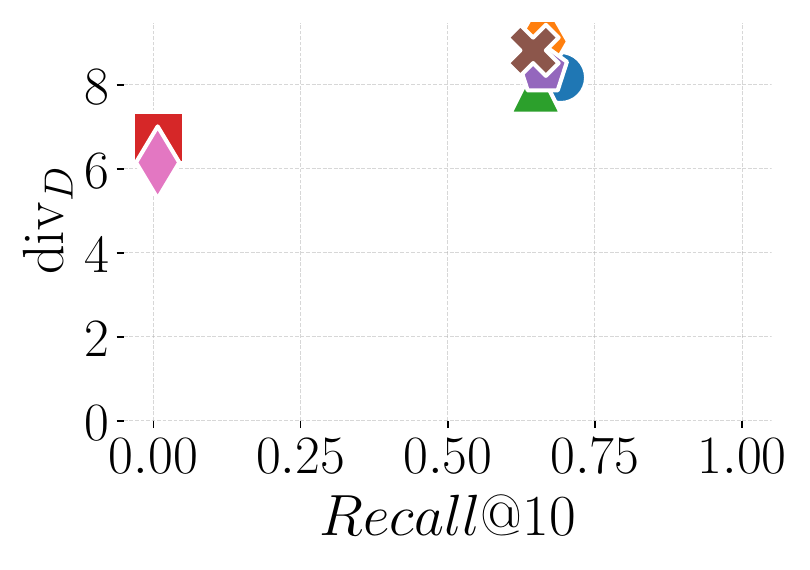}
    \end{subfigure}
    \begin{subfigure}[b]{0.3\textwidth}
        \includegraphics[width=\columnwidth]{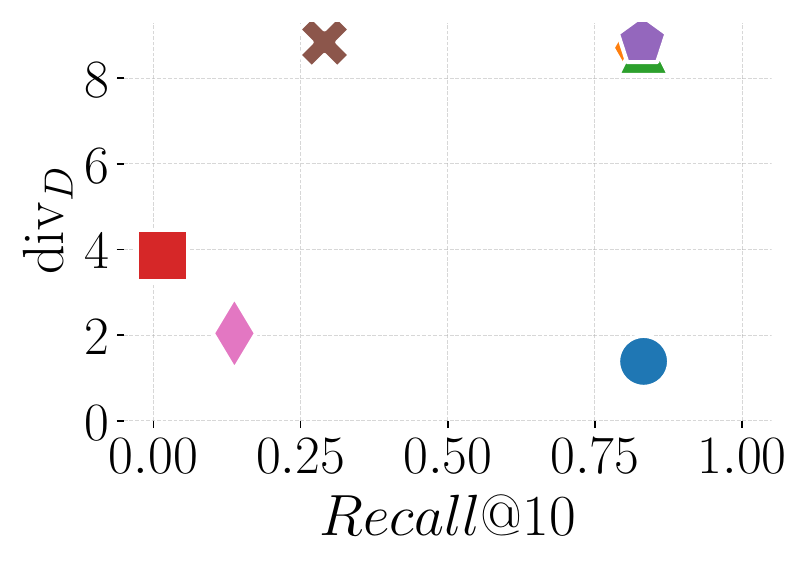}
    \end{subfigure}
    \break
    \begin{subfigure}[b]{0.3\textwidth}
        \includegraphics[width=\columnwidth]{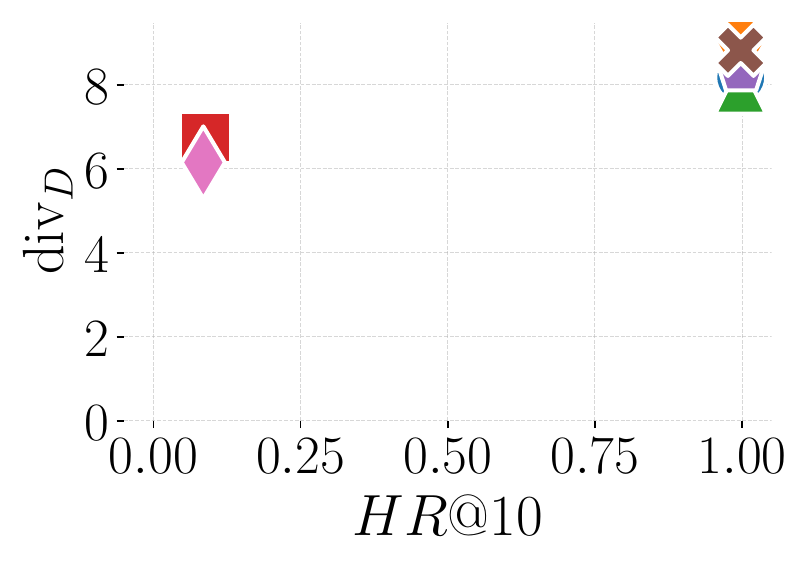}
    \end{subfigure}
    \begin{subfigure}[b]{0.3\textwidth}
        \includegraphics[width=\columnwidth]{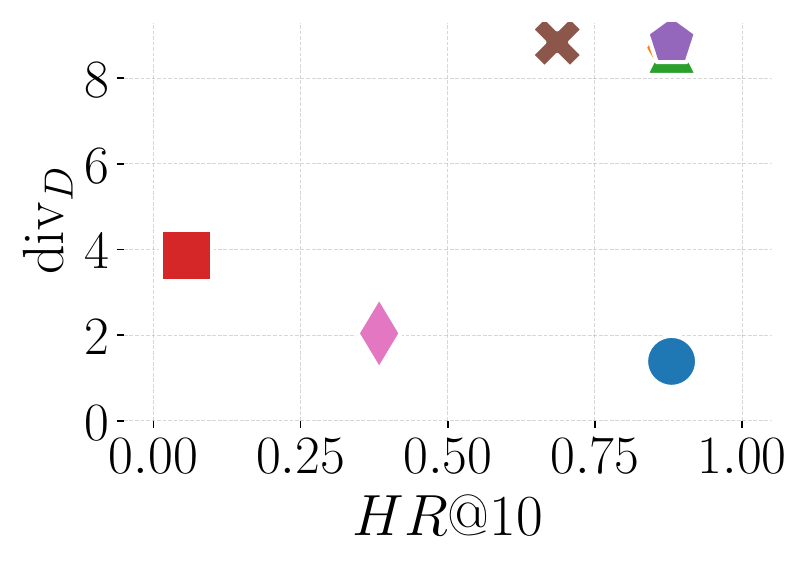}
    \end{subfigure}
    \break
    \begin{subfigure}[b]{0.3\textwidth}
        \includegraphics[width=\columnwidth]{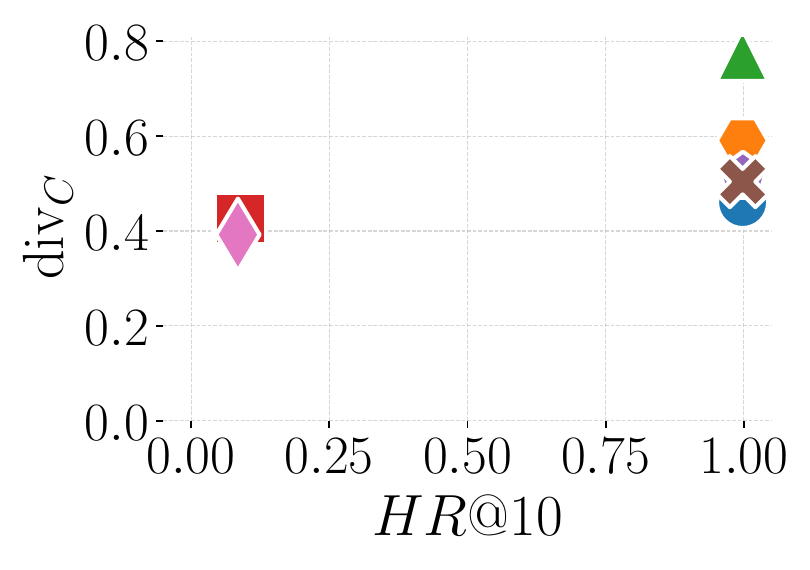}
    % \caption{Movielens-1M}
    \end{subfigure}
    \begin{subfigure}[b]{0.3\textwidth}
        \includegraphics[width=\columnwidth]{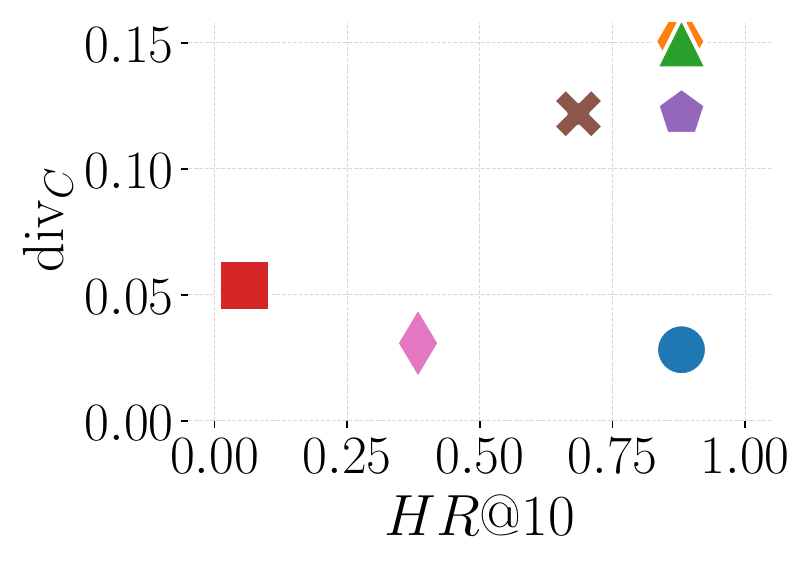}
    \end{subfigure}
    \break
    \begin{subfigure}[b]{0.3\textwidth}
        \includegraphics[width=\columnwidth]{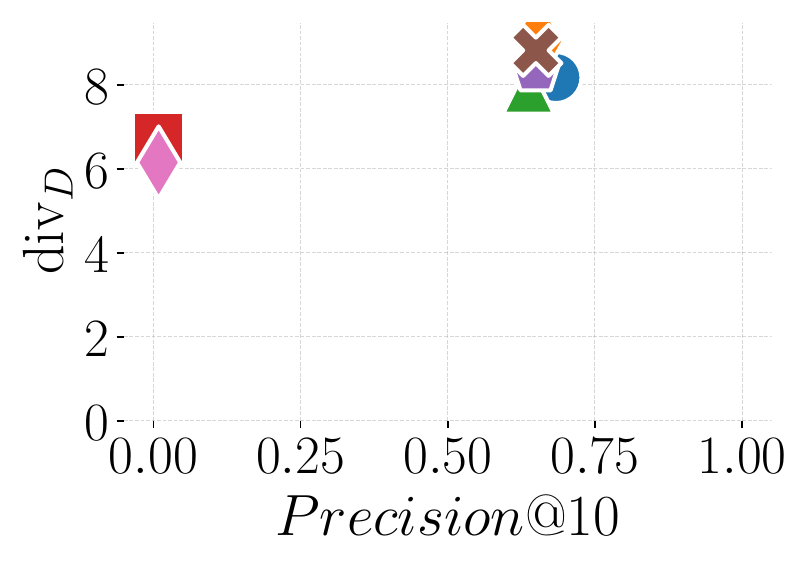}
    \end{subfigure}
    \begin{subfigure}[b]{0.3\textwidth}
        \includegraphics[width=\columnwidth]{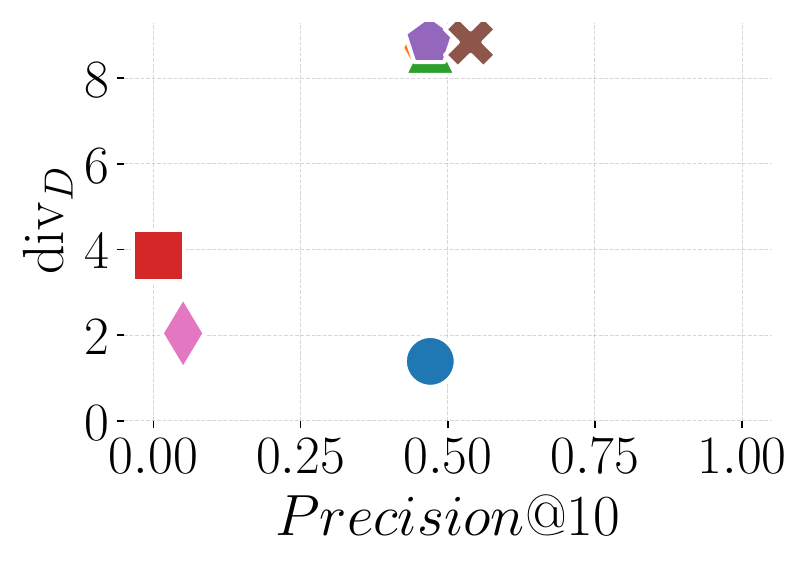}
    \end{subfigure}
    \break
    \begin{subfigure}[b]{0.3\textwidth}
        \includegraphics[width=\columnwidth]{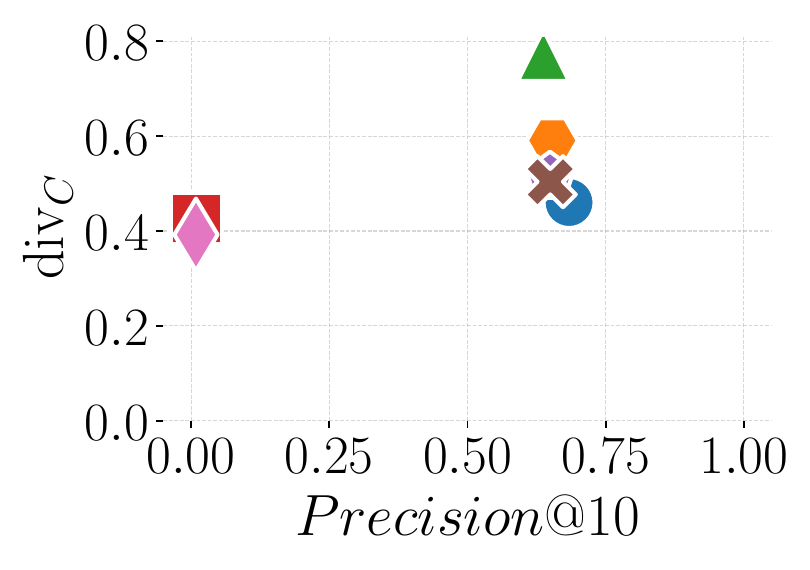}
    \caption{Netflix-Prize}
    \end{subfigure}
    \begin{subfigure}[b]{0.3\textwidth}
        \includegraphics[width=\columnwidth]{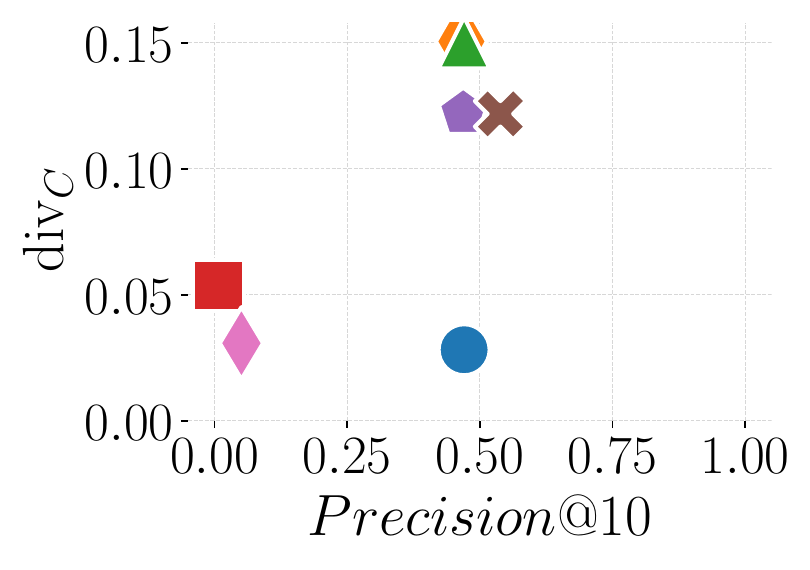}
    \caption{Yahoo-R2}
    \end{subfigure}
    \caption{Trade-off between either \divdist or \divcov and either \rm  \emph{Recall}@10, \emph{HR}@10, \textbf{or} \rm \emph{Precision}@10 \textbf{over Netflix-Prize and Yahoo-R2. The X-axis shows the recommendation quality while the Y-axis represents the diversity score.}}
    \label{fig:quality-diversity-tradeoff-2}
\end{figure*}

Figures~\ref{fig:quality-diversity-tradeoff-1} and~\ref{fig:quality-diversity-tradeoff-2} show the trade-off between recommendation quality and diversity, either in terms of \divdist or \divcov, and either \{\textit{Recall}, \textit{HR}, \textit{Precision}\}@10. Further, Tables~\ref{tab:scores-table-1} and~\ref{tab:scores-table-2} report the scores in terms of \divdist, \divcov, \steps, $\Delta_{\expecteddivdist}$, $\Delta_{\expecteddivcov}$, and $\Delta_{\expectation{\mathrm{steps}}}$, obtained by varying the expected number of steps in the range [5, 10, 20]. 

As we can see, our algorithm, for both variants \ouralgodistance and \ouralgocover, 
outperforms the competitors for most of the datasets, 
offering a significant improvement especially in terms of coverage.

Notably, despite the fact that in some settings \ouralgodistance obtains a slightly lower \divdist score than DPP, it is crucial to consider the timing needed to provide the recommendation list, reported in Figure~\ref{fig:times}. As we can see, competitors such as MMR and DPP require a considerable amount of time to compute their recommended lists, 
in particular for the largest datasets.
Our algorithm, instead, proves to be much more efficient, and its running time is basically constant over all the benchmarks.

\begin{figure*}[t]
\vspace{-2mm}
    \centering
    \begin{subfigure}[b]{0.3\linewidth}
        \includegraphics[width=\columnwidth]{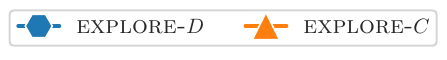}
    \end{subfigure}
    \break
    \begin{subfigure}[b]{\columnwidth}
        \includegraphics[width=\columnwidth]{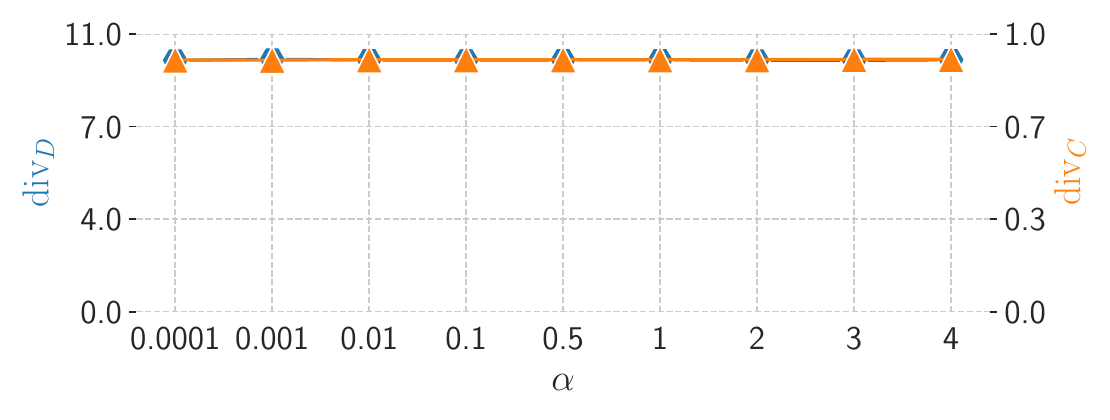}
    \caption{Movielens-1M}
    \end{subfigure}
    \begin{subfigure}[b]{\columnwidth}
        \includegraphics[width=\columnwidth]{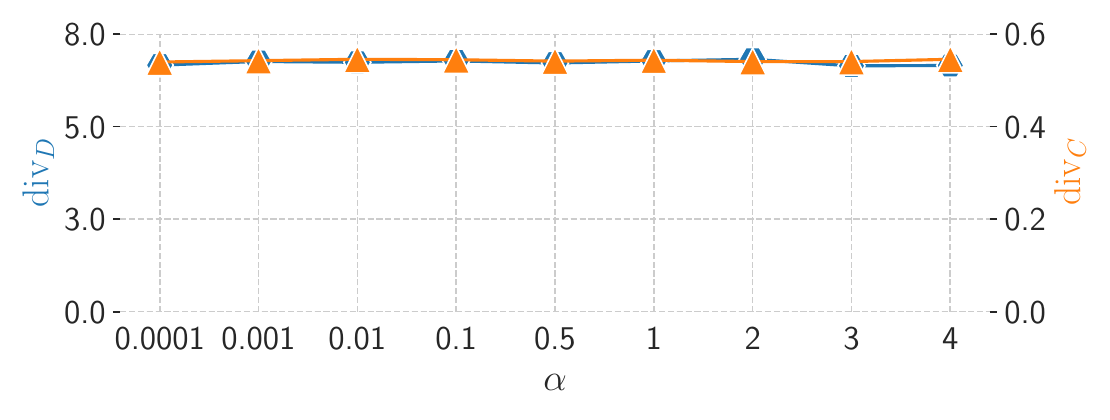}
    \caption{Coat}
    \end{subfigure}
    \begin{subfigure}[b]{\columnwidth}
        \includegraphics[width=\columnwidth]{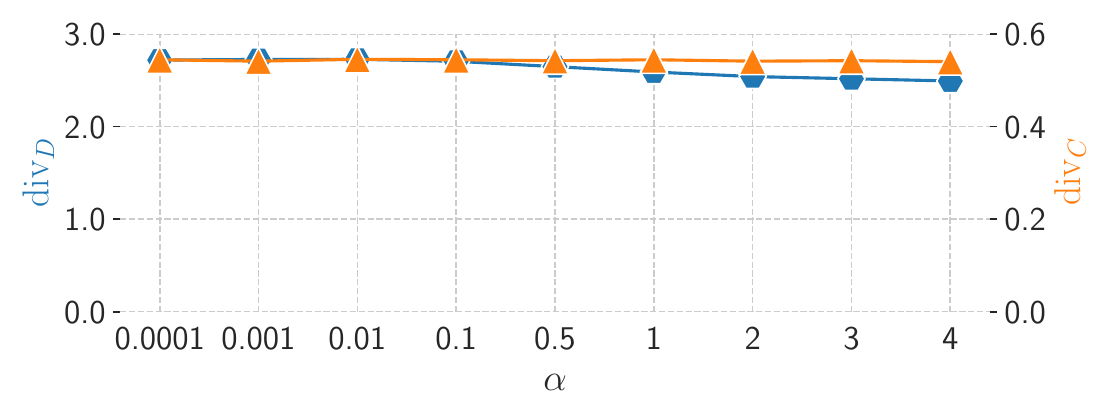}
    \caption{KuaiRec-2.0}
    \end{subfigure}
    \begin{subfigure}[b]{\columnwidth}
        \includegraphics[width=\columnwidth]{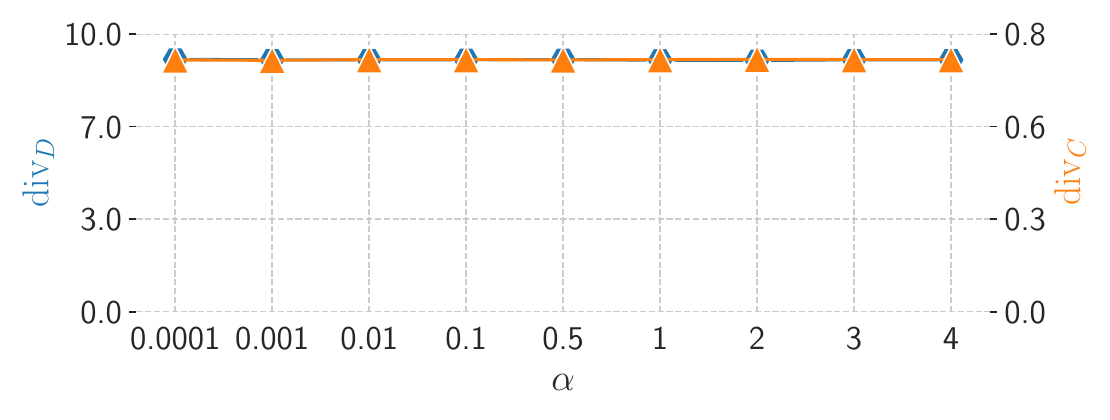}
    \caption{Netflix-Prize}
    \end{subfigure}
    \begin{subfigure}[b]{\columnwidth}
        \includegraphics[width=\columnwidth]{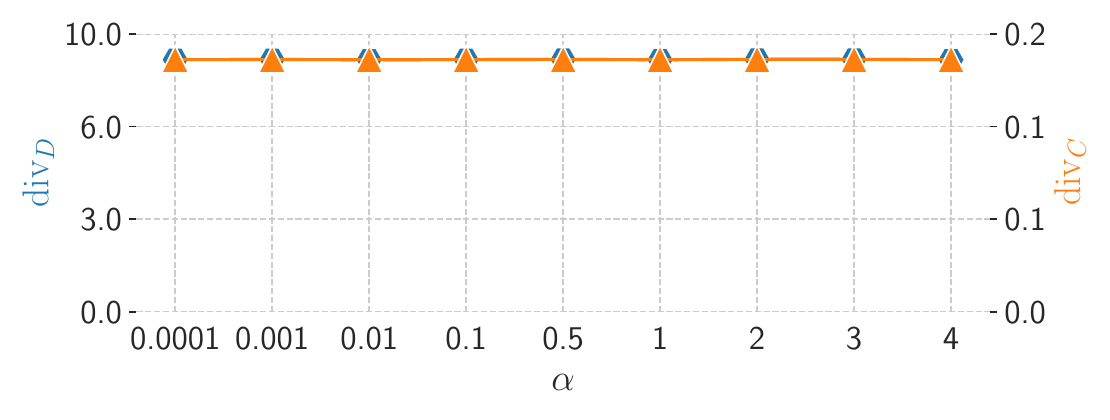}
    \caption{Yahoo-R2}
    \end{subfigure}
    \vspace{-2mm}
    \caption{Effects of tuning the $\alpha$ parameter fixing \expectation{\mathrm{steps}}$ = 10$. The X-axis represents different values of $\alpha$, while the Y-axes report values of \divdist (left) and of \divcov (right).}
    \label{fig:tuning-alpha}
\end{figure*}

In Figure~\ref{fig:tuning-alpha} we show the effect of tuning the $\alpha$ parameter of the Clayton copula function. As we can see, the parameter value does not affect the performance of the algorithm, thus proving \ouralgo to be free of hyper-parameter tuning.

Finally, Table~\ref{tab:weibull_diversity_upper} reports the maximum scores computed either in terms of distance (\expecteddivdist) or coverage (\expecteddivcov), while Table~\ref{tab:notation} summarizes the notation adopted throughout the paper.

\newpage

\begin{table*}[t]
    \caption{Results with $\mathbb{E}[\mathrm{steps}] \in [5, 10, 20]$ over Movielens-1M, Coat and KuaiRec-2.0. Best scores with statistical significance $p < 0.05$ are in bold.}
    \vspace{-3mm}
    \begin{subtable}[b]{\columnwidth}
    \resizebox{\textwidth}{!}{\begin{tabular}{c c c c c c c c}
\toprule
\multicolumn{1}{c}{$\mathbb{E}[\mathrm{steps}]$} & Strategy & \multicolumn{1}{c}{$\mathrm{div}_{\!_D}$} & \multicolumn{1}{c}{$\mathrm{div}_{\!_C}$} & \multicolumn{1}{c}{$\steps$} &\multicolumn{1}{c}{$\Delta_{\expecteddivdist}$} & \multicolumn{1}{c}{$\Delta_{\expecteddivcov}$} &\multicolumn{1}{c}{$\Delta_{\mathbb{E}[\mathrm{steps}]}$ } \\ 
\midrule
\multirow{7}{*}{5} & \multirow{1}{*}{Relevance} & 3.67 & 0.22 & 5.0 & 0.27 &0.73 &0.0 \\ 
\cmidrule(lr){2-8}
& \multirow{1}{*}{\ouralgodistance} & \textbf{4.91} & 0.36 & 4.98 & \textbf{0.03} &0.55 &0.0 \\ 
& \multirow{1}{*}{\ouralgocover} & 4.43 & \textbf{0.71} & 4.71 & 0.12 &\textbf{0.11} &0.06 \\ 
\cmidrule(lr){2-8}
& \multirow{1}{*}{MMR} & 3.96 & 0.29 & 4.57 & 0.22 &0.64 &0.09 \\ 
& \multirow{1}{*}{DUM} & 4.4 & 0.33 & 4.98 & 0.13 &0.59 &0.0 \\ 
& \multirow{1}{*}{DPP} & 4.59 & 0.31 & 4.99 & 0.09 &0.61 &0.0 \\ 
& \multirow{1}{*}{DGREC} & 3.36 & 0.37 & 4.49 & 0.33 &0.54 &0.1 \\ 
\cmidrule(lr){1-8}
\multirow{7}{*}{10} & \multirow{1}{*}{Relevance} & 7.45 & 0.34 & 10.02 & 0.26 &0.64 &-0.0 \\ 
\cmidrule(lr){2-8}
& \multirow{1}{*}{\ouralgodistance} & \textbf{9.77} & 0.63 & 9.85 & \textbf{0.03} &0.32 &0.02 \\ 
& \multirow{1}{*}{\ouralgocover} & 8.84 & \textbf{0.89} & 9.7 & 0.12 &\textbf{0.05} &0.03 \\ 
\cmidrule(lr){2-8}
& \multirow{1}{*}{MMR} & 7.29 & 0.43 & 8.62 & 0.27 &0.54 &0.14 \\ 
& \multirow{1}{*}{DUM} & 8.95 & 0.51 & 10.03 & 0.11 &0.45 &-0.0 \\ 
& \multirow{1}{*}{DPP} & 9.29 & 0.5 & 9.99 & 0.08 &0.46 &0.0 \\ 
& \multirow{1}{*}{DGREC} & 6.89 & 0.49 & 8.92 & 0.31 &0.47 &0.11 \\ 
\cmidrule(lr){1-8}
\multirow{7}{*}{20} & \multirow{1}{*}{Relevance} & 14.96 & 0.48 & 20.04 & 0.25 &0.51 &-0.0 \\ 
\cmidrule(lr){2-8}
& \multirow{1}{*}{\ouralgodistance} & \textbf{18.96} & 0.86 & 19.4 & \textbf{0.05} &0.12 &0.03 \\ 
& \multirow{1}{*}{\ouralgocover} & 16.81 & \textbf{0.97} & 19.76 & 0.16 &\textbf{0.01} &0.01 \\ 
\cmidrule(lr){2-8}
& \multirow{1}{*}{MMR} & 13.45 & 0.57 & 16.49 & 0.33 &0.42 &0.18 \\ 
& \multirow{1}{*}{DUM} & 17.98 & 0.7 & \textbf{20.16} & 0.1 &0.29 &\textbf{-0.01} \\ 
& \multirow{1}{*}{DPP} & 18.6 & 0.71 & 20.02 & 0.07 &0.28 &-0.0 \\ 
& \multirow{1}{*}{DGREC} & 13.91 & 0.63 & 17.64 & 0.31 &0.36 &0.12 \\ 
\bottomrule
\end{tabular}}

    \caption{Movielens-1M}
    \end{subtable}
    \hspace{0.5cm}
    \begin{subtable}[b]{\columnwidth}
    \resizebox{\textwidth}{!}{\begin{tabular}{c c c c c c c c}
\toprule
\multicolumn{1}{c}{$\mathbb{E}[\mathrm{steps}]$} & Strategy & \multicolumn{1}{c}{$\mathrm{div}_{\!_D}$} & \multicolumn{1}{c}{$\mathrm{div}_{\!_C}$} & \multicolumn{1}{c}{$\steps$} &\multicolumn{1}{c}{$\Delta_{\expecteddivdist}$} & \multicolumn{1}{c}{$\Delta_{\expecteddivcov}$} &\multicolumn{1}{c}{$\Delta_{\mathbb{E}[\mathrm{steps}]}$ } \\ 
\midrule
\multirow{7}{*}{5} & \multirow{1}{*}{Relevance} & 3.15 & 0.3 & 4.36 & 0.38 &0.3 &0.13 \\ 
\cmidrule(lr){2-8}
& \multirow{1}{*}{\ouralgodistance} & \textbf{3.48} & 0.34 & 4.16 & \textbf{0.31} &0.2 &0.17 \\ 
& \multirow{1}{*}{\ouralgocover} & 3.36 & \textbf{0.35} & 4.13 & 0.33 &\textbf{0.18} &0.17 \\ 
\cmidrule(lr){2-8}
& \multirow{1}{*}{MMR} & 2.43 & 0.26 & 3.54 & 0.52 &0.39 &0.29 \\ 
& \multirow{1}{*}{DUM} & 3.11 & 0.3 & 4.31 & 0.38 &0.3 &0.14 \\ 
& \multirow{1}{*}{DPP} & 3.28 & 0.31 & 4.33 & 0.35 &0.27 &0.13 \\ 
& \multirow{1}{*}{DGREC} & 2.2 & 0.24 & 3.2 & 0.56 &0.44 &0.36 \\ 
\cmidrule(lr){1-8}
\multirow{7}{*}{10} & \multirow{1}{*}{Relevance} & 6.38 & 0.44 & 8.64 & 0.36 &0.35 &0.14 \\ 
\cmidrule(lr){2-8}
& \multirow{1}{*}{\ouralgodistance} & 6.73 & 0.51 & 8.02 & 0.33 &0.25 &0.2 \\ 
& \multirow{1}{*}{\ouralgocover} & 6.31 & \textbf{0.55} & 7.81 & 0.37 &\textbf{0.19} &0.22 \\ 
\cmidrule(lr){2-8}
& \multirow{1}{*}{MMR} & 4.63 & 0.38 & 6.75 & 0.54 &0.44 &0.32 \\ 
& \multirow{1}{*}{DUM} & 6.44 & 0.46 & 8.65 & 0.36 &0.32 &0.13 \\ 
& \multirow{1}{*}{DPP} & 6.64 & 0.45 & 8.61 & 0.34 &0.34 &0.14 \\ 
& \multirow{1}{*}{DGREC} & 4.56 & 0.37 & 6.33 & 0.55 &0.46 &0.37 \\ 
\cmidrule(lr){1-8}
\multirow{7}{*}{20} & \multirow{1}{*}{Relevance} & 12.23 & 0.59 & 16.36 & 0.39 &0.33 &0.18 \\ 
\cmidrule(lr){2-8}
& \multirow{1}{*}{\ouralgodistance} & 12.85 & 0.7 & 15.48 & 0.36 &0.21 &0.23 \\ 
& \multirow{1}{*}{\ouralgocover} & 12.09 & \textbf{0.77} & 15.36 & 0.4 &\textbf{0.13} &0.23 \\ 
\cmidrule(lr){2-8}
& \multirow{1}{*}{MMR} & 8.79 & 0.53 & 13.12 & 0.56 &0.4 &0.34 \\ 
& \multirow{1}{*}{DUM} & 12.63 & 0.62 & 16.78 & 0.37 &0.3 &0.16 \\ 
& \multirow{1}{*}{DPP} & 13.06 & 0.62 & 16.84 & 0.35 &0.3 &0.16 \\ 
& \multirow{1}{*}{DGREC} & 9.31 & 0.53 & 12.84 & 0.54 &0.4 &0.36 \\ 
\bottomrule
\end{tabular}}

    \caption{Coat}
    \end{subtable}
    \begin{subtable}[b]{\columnwidth}
        \resizebox{\textwidth}{!}{\begin{tabular}{c c c c c c c c}
\toprule
\multicolumn{1}{c}{$\mathbb{E}[\mathrm{steps}]$} & Strategy & \multicolumn{1}{c}{$\mathrm{div}_{\!_D}$} & \multicolumn{1}{c}{$\mathrm{div}_{\!_C}$} & \multicolumn{1}{c}{$\steps$} &\multicolumn{1}{c}{$\Delta_{\expecteddivdist}$} & \multicolumn{1}{c}{$\Delta_{\expecteddivcov}$} &\multicolumn{1}{c}{$\Delta_{\mathbb{E}[\mathrm{steps}]}$ } \\ 
\midrule
\multirow{7}{*}{5} & \multirow{1}{*}{Relevance} & 0.76 & 0.13 & 4.81 & 0.81 &0.74 &0.04 \\ 
\cmidrule(lr){2-8}
& \multirow{1}{*}{\ouralgodistance} & \textbf{1.56} & 0.11 & 3.54 & \textbf{0.61} &0.78 &0.29 \\ 
& \multirow{1}{*}{\ouralgocover} & 1.08 & \textbf{0.34} & 4.09 & 0.73 &\textbf{0.32} &0.18 \\ 
\cmidrule(lr){2-8}
& \multirow{1}{*}{MMR} & 1.25 & 0.12 & 3.89 & 0.68 &0.76 &0.22 \\ 
& \multirow{1}{*}{DUM} & 0.83 & 0.17 & 4.8 & 0.79 &0.66 &0.04 \\ 
& \multirow{1}{*}{DPP} & 1.38 & 0.09 & 4.75 & 0.65 &0.82 &0.05 \\ 
& \multirow{1}{*}{DGREC} & 0.77 & 0.11 & 2.64 & 0.81 &0.78 &0.47 \\ 
\cmidrule(lr){1-8}
\multirow{7}{*}{10} & \multirow{1}{*}{Relevance} & 1.63 & 0.21 & 9.6 & 0.79 &0.71 &0.04 \\ 
\cmidrule(lr){2-8}
& \multirow{1}{*}{\ouralgodistance} & \textbf{3.06} & 0.17 & 6.86 & \textbf{0.61} &0.77 &0.31 \\ 
& \multirow{1}{*}{\ouralgocover} & 2.22 & \textbf{0.53} & 7.95 & 0.72 &\textbf{0.28} &0.2 \\ 
\cmidrule(lr){2-8}
& \multirow{1}{*}{MMR} & 2.52 & 0.2 & 7.34 & 0.68 &0.73 &0.27 \\ 
& \multirow{1}{*}{DUM} & 1.78 & 0.27 & 9.59 & 0.77 &0.63 &0.04 \\ 
& \multirow{1}{*}{DPP} & 2.81 & 0.15 & 9.54 & 0.64 &0.8 &0.05 \\ 
& \multirow{1}{*}{DGREC} & 1.71 & 0.19 & 5.45 & 0.78 &0.74 &0.45 \\ 
\cmidrule(lr){1-8}
\multirow{7}{*}{20} & \multirow{1}{*}{Relevance} & 3.62 & 0.32 & 19.02 & 0.77 &0.65 &0.05 \\ 
\cmidrule(lr){2-8}
& \multirow{1}{*}{\ouralgodistance} & \textbf{6.16} & 0.26 & 13.83 & \textbf{0.61} &0.71 &0.31 \\ 
& \multirow{1}{*}{\ouralgocover} & 4.55 & \textbf{0.76} & 15.09 & 0.71 &\textbf{0.16} &0.25 \\ 
\cmidrule(lr){2-8}
& \multirow{1}{*}{MMR} & 4.79 & 0.3 & 13.91 & 0.69 &0.67 &0.3 \\ 
& \multirow{1}{*}{DUM} & 3.86 & 0.39 & 19.09 & 0.75 &0.57 &0.05 \\ 
& \multirow{1}{*}{DPP} & 5.56 & 0.24 & 18.99 & 0.64 &0.73 &0.05 \\ 
& \multirow{1}{*}{DGREC} & 3.55 & 0.3 & 11.33 & 0.77 &0.67 &0.43 \\ 
\bottomrule
\end{tabular}}

    \caption{KuaiRec-2.0}    
    \end{subtable}
    \hspace{0.5cm}
    \begin{subtable}[b]{\columnwidth}
        \resizebox{\textwidth}{!}{\begin{tabular}{c c c c c c c c}
\toprule
\multicolumn{1}{c}{$\mathbb{E}[\mathrm{steps}]$} & Strategy & \multicolumn{1}{c}{$\mathrm{div}_{\!_D}$} & \multicolumn{1}{c}{$\mathrm{div}_{\!_C}$} & \multicolumn{1}{c}{$\steps$} &\multicolumn{1}{c}{$\Delta_{\expecteddivdist}$} & \multicolumn{1}{c}{$\Delta_{\expecteddivcov}$} &\multicolumn{1}{c}{$\Delta_{\mathbb{E}[\mathrm{steps}]}$ } \\ 
\midrule
\multirow{7}{*}{5} & \multirow{1}{*}{Relevance} & 4.04 & 0.32 & 4.86 & 0.2 &0.59 &0.03 \\ 
\cmidrule(lr){2-8}
& \multirow{1}{*}{\ouralgodistance} & \textbf{4.62} & 0.38 & 4.75 & \textbf{0.09} &0.51 &0.05 \\ 
& \multirow{1}{*}{\ouralgocover} & 3.97 & \textbf{0.6} & 4.43 & 0.21 &\textbf{0.22} &0.11 \\ 
\cmidrule(lr){2-8}
& \multirow{1}{*}{MMR} & 3.56 & 0.3 & 4.19 & 0.3 &0.61 &0.16 \\ 
& \multirow{1}{*}{DUM} & 4.16 & 0.36 & 4.89 & 0.18 &0.53 &0.02 \\ 
& \multirow{1}{*}{DPP} & 4.38 & 0.34 & 4.88 & 0.13 &0.56 &0.02 \\ 
& \multirow{1}{*}{DGREC} & 3.0 & 0.26 & 3.74 & 0.41 &0.66 &0.25 \\ 
\cmidrule(lr){1-8}
\multirow{7}{*}{10} & \multirow{1}{*}{Relevance} & 8.17 & 0.46 & 9.73 & 0.19 &0.47 &0.03 \\ 
\cmidrule(lr){2-8}
& \multirow{1}{*}{\ouralgodistance} & \textbf{9.03} & 0.59 & 9.24 & \textbf{0.1} &0.32 &0.08 \\ 
& \multirow{1}{*}{\ouralgocover} & 7.92 & \textbf{0.77} & 8.69 & 0.21 &\textbf{0.12} &0.13 \\ 
\cmidrule(lr){2-8}
& \multirow{1}{*}{MMR} & 6.76 & 0.43 & 7.94 & 0.33 &0.51 &0.21 \\ 
& \multirow{1}{*}{DUM} & 8.36 & 0.51 & 9.72 & 0.17 &0.41 &0.03 \\ 
& \multirow{1}{*}{DPP} & 8.82 & 0.5 & 9.73 & 0.12 &0.43 &0.03 \\ 
& \multirow{1}{*}{DGREC} & 6.15 & 0.39 & 7.44 & 0.39 &0.55 &0.26 \\ 
\cmidrule(lr){1-8}
\multirow{7}{*}{20} & \multirow{1}{*}{Relevance} & 16.19 & 0.6 & 19.29 & 0.19 &0.34 &0.04 \\ 
\cmidrule(lr){2-8}
& \multirow{1}{*}{\ouralgodistance} & 17.38 & 0.77 & 17.98 & \textbf{0.13} &0.15 &0.1 \\ 
& \multirow{1}{*}{\ouralgocover} & 15.6 & \textbf{0.87} & 17.53 & 0.22 &\textbf{0.04} &0.12 \\ 
\cmidrule(lr){2-8}
& \multirow{1}{*}{MMR} & 12.79 & 0.56 & 15.34 & 0.36 &0.38 &0.23 \\ 
& \multirow{1}{*}{DUM} & 16.59 & 0.65 & 19.33 & 0.17 &0.29 &0.03 \\ 
& \multirow{1}{*}{DPP} & \textbf{17.49} & 0.66 & 19.36 & \textbf{0.13} &0.27 &0.03 \\ 
& \multirow{1}{*}{DGREC} & 12.11 & 0.53 & 14.59 & 0.4 &0.42 &0.27 \\ 
\bottomrule
\end{tabular}}

    \caption{Netflix-Prize}
    \end{subtable}
    % \break
    % \begin{subtable}[b]{\columnwidth}
    %  \input{yahoo-r2/weibull/scores_table}
    % \caption{Yahoo-R2}
    % \end{subtable}
    \label{tab:scores-table-1}
\end{table*}

\begin{table*}[t]
    \caption{Results with $\mathbb{E}[\mathrm{steps}] \in [5, 10, 20]$ over Yahoo-R2. Best scores with statistical significance $p < 0.05$ are in bold.}
    \centering
    \vspace{-3mm}
    \begin{subtable}[b]{\columnwidth}
     \resizebox{\textwidth}{!}{\begin{tabular}{c c c c c c c c}
\toprule
\multicolumn{1}{c}{$\mathbb{E}[\mathrm{steps}]$} & Strategy & \multicolumn{1}{c}{$\mathrm{div}_{\!_D}$} & \multicolumn{1}{c}{$\mathrm{div}_{\!_C}$} & \multicolumn{1}{c}{$\steps$} &\multicolumn{1}{c}{$\Delta_{\expecteddivdist}$} & \multicolumn{1}{c}{$\Delta_{\expecteddivcov}$} &\multicolumn{1}{c}{$\Delta_{\mathbb{E}[\mathrm{steps}]}$ } \\ 
\midrule
\multirow{7}{*}{5} & \multirow{1}{*}{Relevance} & 0.66 & 0.02 & \textbf{4.77} & 0.87 &0.77 &\textbf{0.05} \\ 
\cmidrule(lr){2-8}
& \multirow{1}{*}{\ouralgodistance} & \textbf{4.4} & \textbf{0.08} & 4.49 & \textbf{0.13} &\textbf{0.1} &0.1 \\ 
& \multirow{1}{*}{\ouralgocover} & 4.38 & \textbf{0.08} & 4.47 & \textbf{0.13} &\textbf{0.1} &0.11 \\ 
\cmidrule(lr){2-8}
& \multirow{1}{*}{MMR} & 2.45 & 0.04 & 3.96 & 0.52 &0.55 &0.21 \\ 
& \multirow{1}{*}{DUM} & 4.38 & 0.07 & 4.72 & \textbf{0.13} &0.21 &0.06 \\ 
& \multirow{1}{*}{DPP} & 4.38 & 0.07 & 4.72 & \textbf{0.13} &0.21 &0.06 \\ 
& \multirow{1}{*}{DGREC} & 1.02 & 0.02 & 3.68 & 0.8 &0.77 &0.26 \\ 
\cmidrule(lr){1-8}
\multirow{7}{*}{10} & \multirow{1}{*}{Relevance} & 1.39 & 0.03 & \textbf{9.49} & 0.86 &0.83 &\textbf{0.05} \\ 
\cmidrule(lr){2-8}
& \multirow{1}{*}{\ouralgodistance} & 8.71 & \textbf{0.15} & 8.73 & 0.13 &\textbf{0.14} &0.13 \\ 
& \multirow{1}{*}{\ouralgocover} & 8.67 & \textbf{0.15} & 8.7 & 0.14 &\textbf{0.14} &0.13 \\ 
\cmidrule(lr){2-8}
& \multirow{1}{*}{MMR} & 3.86 & 0.05 & 7.36 & 0.62 &0.71 &0.26 \\ 
& \multirow{1}{*}{DUM} & 8.86 & 0.12 & 9.43 & 0.12 &0.31 &0.06 \\ 
& \multirow{1}{*}{DPP} & 8.84 & 0.12 & 9.41 & 0.12 &0.31 &0.06 \\ 
& \multirow{1}{*}{DGREC} & 2.04 & 0.03 & 7.35 & 0.8 &0.83 &0.26 \\ 
\cmidrule(lr){1-8}
\multirow{7}{*}{20} & \multirow{1}{*}{Relevance} & 3.0 & 0.04 & \textbf{18.81} & 0.85 &0.88 &\textbf{0.06} \\ 
\cmidrule(lr){2-8}
& \multirow{1}{*}{\ouralgodistance} & 16.48 & \textbf{0.28} & 16.5 & 0.18 &\textbf{0.17} &0.18 \\ 
& \multirow{1}{*}{\ouralgocover} & 16.43 & \textbf{0.28} & 16.46 & 0.18 &\textbf{0.17} &0.18 \\ 
\cmidrule(lr){2-8}
& \multirow{1}{*}{MMR} & 5.9 & 0.07 & 13.89 & 0.71 &0.79 &0.31 \\ 
& \multirow{1}{*}{DUM} & 17.59 & 0.19 & 18.73 & 0.12 &0.44 &\textbf{0.06} \\ 
& \multirow{1}{*}{DPP} & 17.6 & 0.19 & 18.74 & 0.12 &0.44 &\textbf{0.06} \\ 
& \multirow{1}{*}{DGREC} & 3.92 & 0.04 & 14.56 & 0.8 &0.88 &0.27 \\ 
\bottomrule
\end{tabular}}

    \caption{Yahoo-R2}
    \end{subtable}
\label{tab:scores-table-2}
\end{table*}

% \vspace{-2mm}
\newpage

\begin{table*}[t]
    \centering
    \caption{Maximum scores in terms of diversity and coverage per dataset, by varying the expected number of steps.}
    \vspace{-2mm}  
    \begin{tabular}{l r r r}
    \toprule
         Dataset & $\expectation{\mathrm{steps}}$ & $\expecteddivdist$ & $\expecteddivcov$ \\
         \midrule
         \multirow{3}{*}{Movielens-1M} & 5 & 5.05 & 0.80 \\ & 10 & 10.05 & 0.93 \\ & 20 & 20.03 & 0.98 \\
         \cmidrule(lr){1-4}
         \multirow{3}{*}{Coat} & 5 & 5.05 & 0.43 \\ & 10 & 10.05 & 0.68 \\ & 20 & 20.03 & 0.89 \\
         \cmidrule(lr){1-4}
         \multirow{3}{*}{KuaiRec-2.0} & 5 & 3.97 & 0.50 \\ & 10 & 7.87 & 0.73 \\ & 20 & 15.63 & 0.91 \\
         \cmidrule(lr){1-4}
         \multirow{3}{*}{Netflix-Prize} & 5 & 5.05 & 0.77 \\ & 10 & 10.05 & 0.87 \\ & 20 & 20.03 & 0.91 \\
         \cmidrule(lr){1-4}
         \multirow{3}{*}{Yahoo-r2} & 5 & 5.05 & 0.09 \\ & 10 & 10.05 & 0.17 \\ & 20 & 20.03 & 0.34 \\
    \bottomrule
    \label{tab:weibull_diversity_upper}
    \end{tabular}
\end{table*}

\begin{table*}[t]
    \centering
    \caption{Notation}
    \begin{tabular}{l l}
    \toprule
      \users & Set of users \\
      \items & Set of items \\
      \categories &  Set of categories \\
      $\mathcal{R}$(\auser, \anitem) & Relevance score of the item \anitem for the user \auser \\
      \strategy & Recommendation strategy \\
      \vectxi & Users vector of item \anitem \\
      \vectyi &  Categories vector of item \anitem \\
      $d$(\anitem, \anotheritem) & Weighted Jaccard distance between two items \anitem, \anotheritem, either in terms of users or categories \\
      \divdist & Diversity score in terms of distance \\
      \divcov & Diversity score in terms of coverage \\
      \ilistk & Recommendation list produced at step $t$ \\
      \ilistsize & Size of the recommendation lists \\
      \bufferprobi & Probability for the user to be interested in the item \anitem \\
      \selectprobi & Probability for the user to select the item \anitem \\
      \quitprobk & Weariness probability, i.e., probability for the user to lose interest at step $t$ \\
      \quitatlistk & Quitting probability, i.e., probability to quit the exploration process at step $t$ \\
      \totalquit & Overall quitting probability, i.e., probability to quit the exploration process at any step \\
      \seenitems & Final user interactions set \\
      \steps & Actual number of steps performed by the user at the end of the exploration process \\
      $\expectation{\mathrm{steps}}$ & Expected number of steps to be performed by the user at the end of the exploration process \\
    \bottomrule
    \end{tabular}
    \label{tab:notation}
\end{table*}

\end{document}